# Tuning Elastic Waves in Soft Phononic Crystal Cylinders via Large Deformation and Electromechanical Coupling

Bin Wu[1], Weijian Zhou[1], Ronghao Bao[1], and Weiqiu Chen[1,2,3,4,*]

[1] Department of Engineering Mechanics, Zhejiang University, Hangzhou 310027, China

[2] State Key Laboratory of Fluid Power and Mechatronic Systems, Zhejiang University, Hangzhou 310027, China

[3] Key Laboratory of Soft Machines and Smart Devices of Zhejiang Province, Zhejiang University, Hangzhou 310027, P.R. China

[4] Soft Matter Research Center, Zhejiang University, Hangzhou 310027, P.R. China

**Abstract**

Soft electroactive materials can undergo large deformation subjected to either mechanical or electrical stimulus, and hence they can be excellent candidates for designing extremely flexible and adaptive structures and devices. This paper proposes a simple one-dimensional soft phononic crystal (PC) cylinder made of dielectric elastomer to show how large deformation and electric field can be used jointly to tune the longitudinal waves propagating in the PC. A series of soft electrodes, which are mechanically negligible, are placed periodically along the dielectric elastomer cylinder, and hence the material can be regarded as uniform in the undeformed state. This is also the case for the uniformly pre-stretched state induced by a static axial force only. The effective periodicity of the structure is then achieved through two loading paths, i.e. by maintaining the longitudinal stretch and applying an electric voltage over any two neighbouring electrodes, or by holding the axial force and applying the voltage. All physical field variables for both configurations can be determined exactly based on the nonlinear theory of electroelasticity. An infinitesimal wave motion is further superimposed on the pre-deformed configurations and the corresponding dispersion equations are derived analytically by invoking the linearized theory for incremental motions. Numerical examples are finally considered to show the tunability of wave propagation behavior in the soft PC cylinder. The outstanding performance regarding the band gap (BG) property of the proposed soft dielectric PC is clearly demonstrated by comparing with the conventional design adopting the hard piezoelectric material. One particular point that should be emphasized is that soft dielectric PCs are susceptible to various kinds of failure (buckling, electromechanical instability, electric breakdown, etc.), imposing corresponding limits on the external stimuli. This has been carefully examined for the present soft PC cylinder such that the applied electric voltage is always assumed to be less than the critical voltage except for one case, in which we illustrate that the snap-through instability of the axially free PC cylinder made of a generalized Gent material may be used to efficiently trigger a sharp transition in the BGs.

[*] Corresponding author. Email: chenwq@zju.edu.cn; Tel./Fax: 86-571-87951866.

1. Introduction

Phononic crystals (PCs) have attracted intensive research interest for decades since the early 1990s. They exhibit a spatially periodic feature regarding the distribution of physical properties, geometry, boundary conditions, or/and external stimuli, making their wave spectrum or band structure quite different from that of the conventional homogeneous media. A prominent feature is that PCs are endowed with the property of frequency band gaps (BGs) wherein waves are prohibited to propagate, i.e. they will be attenuated rapidly during propagation. The BGs are formed due to Bragg scattering and can be utilized to manipulate waves that are the basis of many novel wave devices. Another mechanism to form BGs is local resonance, and the materials with units of local resonance are now known as metamaterials (MMs). For the recent progress in PCs and MMs, the reader is referred to, for instance, the excellent book edited by Deymier [1] and the comprehensive review articles by Hussein et al. [2], Ma and Sheng [3], and Cummer et al. [4].

Once a PC is manufactured, its band structure in general is fixed. However, to be able to control waves in materials at will is of practical importance in many areas such as smart wave devices, seismology, vibration isolators, energy harvesters, etc. Goffaux and Vignerons [5] first put forward the concept of tunable PC and suggested to rotate the solid scatters in an air/solid phononic system to actively tune the band structure. Since then, tunable PCs have become a main subject in the area, and diverse ways have been developed, indicating a brilliant prospect in optional manipulation of waves via PCs (see Chapter 8 of Deymier [1] for instance). A particularly attractive approach is to exploit the reversible large deformation capability of soft elastic materials. Large deformation changes the structural geometry as well as the material effective stiffness, and further alters the wave propagation behavior. The study along this line dates back to the seminal work of Hayes and Rivlin [6], and more historical background may be found in the book edited by Destrade and Sacoomandi [7]. However, it has been noticed only in the recent decade in the study of PCs. Parnell [8] first carried out an analysis of how the initial finite elastic deformation of a periodic composite bar affects the location of BGs. A big breakthrough immediately follows that Bertoldi and Boyce [9] suggested to make use of instability and post-buckling deformation to tune waves propagating in elastomeric PCs. This scheme opens avenues for a wide range of applications of tunable PCs [10-16]. There are also a lot of parallel studies that directly employ pre-buckling large deformations in a theoretical sense to control waves/vibrations in various PCs [17-23].

Another interesting idea is to explore the application of multi-field coupling effects in soft PCs. The first piece of work seems to appear in 2007 by Yang and Chen [24], who considered a two-dimensional (2D) PC consisting of a square array of parallel dielectric elastomer (DE) cylindrical actuators in air and demonstrated that the acoustic BG could be conveniently tuned by means of the electric field. Later they showed that the refractive and focusing behaviors of a sonic crystals with DE cylindrical actuators also could be adjusted electrically [25]. Various optical and acoustic devices incorporating DE components with electrically tunable functions were further proposed and studied [26-28]. Gei et al. [29] carried out an independent study by considering a thin electrically actuated pre-stretched DE, which is able to filter waves in the assigned ranges of frequencies. Shmuel and deBotton [30] studied the electrostatically controllable BGs of incremental thickness-shear waves in ideal neo-Hookean DE laminates by the transfer matrix method and they only considered the specific case of shear waves propagating perpendicular to the DE layers. Shmuel and Pernas-Salomón [31] proposed an effective way to realize mode localization in DE films via electrostatically controlled aperiodicity. Noticing that the analysis in Yang and Chen [24] is not fully nonlinear, Bortot and Shmuel [32] performed a re-examination and found that the snap-through instability resulting from geometrical and material nonlinearities can be harnessed to achieve sharp transitions in the acoustic BG. Galich and Rudykh [33] re-examined the problem that was studied by Shmuel and deBotton [30] and arrived at a different conclusion that shear waves propagating perpendicular to the layers in ideal neo-Hookean DE laminates are not affected by the electric field directly. In addition, Galich and Rudykh [33] derived long wave estimates for phase and group velocities of the shear waves propagating in any direction. Recently, utilizing the supercell plane wave expansion method, Getz and Shmuel [34] investigated the BG tunability in incompressible DE fiber composite plates and pointed out the application of soft dielectric films in active wave manipulators. Besides the electromechanical coupling, the magnetorheological and magnetoelastic characteristics have also been employed to actively tune the waves in soft PCs [35, 36].

Inspired by the earlier works mentioned above, we have conducted a systematic investigation in recent years on wave propagation in soft materials and structures in various configurations [17, 18, 37-

42]. In this paper, we intend to show that the combination of large deformation and electromechanical coupling can lead to a very flexible and efficient way to tune the BGs by considering a simple phononic cylinder of soft electroactive elastomer with periodic electric boundary conditions. The PC cylinder actually consists of identical homogeneous sub-cylinders of finite length with mechanically negligible electrodes bonded onto their end surfaces. These sub-cylinders are connected by a small amount of insulating glue (such as epoxy), which is also mechanically negligible. A large elastic deformation is first induced by an axial force to form a uniformly pre-stretched configuration. Then we consider two loading paths to form the effective periodicity, and to arrive at the desired initial configuration of the soft PC cylinder. For Path A, we keep the longitudinal stretch unchanged and apply an identical electric voltage difference to each sub-cylinder. For Path B, we keep the axial force unchanged and apply the electric voltage. These two paths give birth to two macroscopically different PCs so that we can study the effect of the formative paths on the BGs. Based on the nonlinear theory of electroelasticity proposed by Dorfmann and Ogden [43], we are able to derive exactly all nonlinear field variables in the initial configuration without any approximation. Then we use an analytical method to obtain the explicit dispersion equations for infinitesimal incremental wave motions that are further superimposed on the initial configuration [44]. The only approximation made in the derivation is the one-dimensional stress state which is well-accepted in the classical rod theory. Analytical expression of the critical voltage resulting in zero-frequency of the first BG and zero effective wave velocity is obtained for both loading paths, which defines an upper boundary of the allowable applied voltage. Based on the derived dispersion equations, we numerically calculate the band structures of these two different PC cylinders that are formed through Paths A and B respectively. In particular, the influence of strain-stiffening effect of the nonlinear material model is addressed by studying how the snap-through instability of a soft PC made of the generalized Gent material can be used to tune the BGs.

The paper is outlined as follows. Section 2 is devoted to a brief description of basic formulations for the nonlinear theory of electroelasticity and the associated linearized theory of incremental wave motions. The uniform nonlinear response of a single homogeneous DE cylinder to an axial force and an axial electric voltage is analyzed in detail for the ideal compressible Gent model in Section 3. Section 4 determines exactly the nonlinear electroelastic fields for the Gent model in the two initial configurations (for Paths A and B, respectively), while in Section 5 we carry out an approximate but analytical analysis of the superimposed linear waves in the soft PCs. Numerical examples are considered in Section 6, and some concluding remarks are made in Section 7. The analytical results for the ideal compressible neo-Hookean model that are useful for conducting numerical calculation are explicitly given in the appendix.

## 2. Theoretical Background

To describe the kinematics of a continuous electroelastic body, we denote the position vector of a particle $X$ in the undeformed state (the reference configuration) as $\mathbf{X}$, which moves to a new position $\mathbf{x}$ in the deformed state (the current configuration). The mapping $\mathbf{x} = \chi(\mathbf{X},t)$, with $t$ being the time variable, is assumed to be sufficiently smooth, and its derivative with respect to $\mathbf{X}$ defines the deformation gradient tensor $\mathbf{F} = \text{Grad}\,\chi = \partial \mathbf{x}/\partial \mathbf{X}$. Then, the local measure of change in material volume is given by $J = \det \mathbf{F}$, which always equals one for an incompressible material. Nonetheless, the material compressibility must be considered in this work to investigate the longitudinal wave propagation. In the following, we will adopt the general theoretical framework of nonlinear electroelasticity suggested by Dorfmann and Ogden [43, 44], but we also note that different versions of the theory should be all equivalent [45].

### 2.1 Basic Formulations for Nonlinear Electroelasticity

We express the basic formulations in Lagrangian form, which can be easily transformed into the corresponding Eulerian form. The nonlinear constitutive relations for a compressible DE read as

$$\mathbf{S} = \frac{\partial W(\mathbf{F},\mathbf{D}_L)}{\partial \mathbf{F}}, \quad \mathbf{E}_L = \frac{\partial W(\mathbf{F},\mathbf{D}_L)}{\partial \mathbf{D}_L} \tag{1}$$

where $\mathbf{S} = J\mathbf{F}^{-1}\boldsymbol{\sigma}$ is the total nominal stress tensor with $\boldsymbol{\sigma}$ being the total Cauchy stress tensor, $\mathbf{E}_L = \mathbf{F}^T\mathbf{E}$ and $\mathbf{D}_L = J\mathbf{F}^{-1}\mathbf{D}$ are the Lagrangian counterparts of the current electric field vector $\mathbf{E}$ and the electric displacement vector $\mathbf{D}$, respectively. Note that $W(\mathbf{F},\mathbf{D}_L)$ is the energy density function per unit volume in the reference configuration. The electric field is governed by the following equations for electrostatics:

$$\text{Curl}\,\mathbf{E}_L = \mathbf{0}, \quad \text{Div}\,\mathbf{D}_L = 0 \tag{2}$$

Here and in the following, the differential operators with the first upper-case letter correspond to the reference configuration, while those with the first lower-case letter refer to the current configuration. The mechanical equations of motion are

$$\text{Div}\,\mathbf{S} = \rho_r \mathbf{x}_{,tt} \tag{3}$$

where the subscript $t$ following a comma denotes the material time derivative, $\rho_r$ is the mass density in the reference configuration, and it is related to the current mass density $\rho$ by the relation $\rho_r = J\rho$.

Equations (1)-(3) plus appropriate initial and boundary conditions, which will be specified later in Sections 3 and 4 for the problem studied, will fully determine the nonlinear response of the electroelastic body subjected to the external mechanical and electrical stimuli. It is noted that the electric field in the vacuum is assumed to have little effect on the motion of the electroelastic body, and is neglected in the following analysis.

**2.2 Linearized Theory for an Incremental Motion**

If an infinitesimal incremental wave motion is superimposed on a finitely deformed electroelastic body, which is in the initial configuration associated with the mapping $\mathbf{x} = \chi(\mathbf{X})$, then we have the following equations for the incremental fields in Eulerian or updated Lagrangian form:

$$\mathrm{div}\,\dot{\mathbf{S}}_0 = \rho \mathbf{u}_{,tt} \tag{4}$$

$$\mathrm{curl}\,\dot{\mathbf{E}}_{L0} = \mathbf{0}, \quad \mathrm{div}\,\dot{\mathbf{D}}_{L0} = 0 \tag{5}$$

$$\dot{\mathbf{S}}_0 = \mathcal{A}_0 \mathbf{H} + \mathcal{B}_0 \dot{\mathbf{D}}_{L0}, \quad \dot{\mathbf{E}}_{L0} = \mathcal{B}_0^{\mathrm{T}} \mathbf{H} + \mathcal{C}_0 \dot{\mathbf{D}}_{L0} \tag{6}$$

where a superposed dot indicates an increment, $\mathbf{u}(\mathbf{x},t) = \dot{\mathbf{x}}(\mathbf{X},t)$, $\dot{\mathbf{S}}_0 = J^{-1}\mathbf{F}\dot{\mathbf{S}}$, $\dot{\mathbf{E}}_{L0} = \mathbf{F}^{-\mathrm{T}}\dot{\mathbf{E}}_L$ and $\dot{\mathbf{D}}_{L0} = J^{-1}\mathbf{F}\dot{\mathbf{D}}_L$ are the 'push forward' quantities, all taking the deformed configuration as the new reference configuration, $\mathbf{H} = \mathrm{grad}\,\mathbf{u}$ is the incremental displacement gradient, and the instantaneous electroelastic tensors $\mathcal{A}_0$, $\mathcal{B}_0$ and $\mathcal{C}_0$ are defined as

$$\mathcal{A}_{0piqj} = J^{-1}F_{p\alpha}F_{q\beta}\mathcal{A}_{\alpha i \beta j} = \mathcal{A}_{0qjpi}, \quad \mathcal{B}_{0piq} = F_{p\alpha}F_{\beta q}^{-1}\mathcal{B}_{\alpha i \beta} = \mathcal{B}_{0ipq}, \quad \mathcal{C}_{0ij} = JF_{\alpha i}^{-1}F_{\beta j}^{-1}\mathcal{C}_{\alpha\beta} = \mathcal{C}_{0ji} \tag{7}$$

with

$$\mathcal{A} = \frac{\partial^2 W}{\partial \mathbf{F} \partial \mathbf{F}}, \quad \mathcal{B} = \frac{\partial^2 W}{\partial \mathbf{F} \partial \mathbf{D}_L}, \quad \mathcal{C} = \frac{\partial^2 W}{\partial \mathbf{D}_L \partial \mathbf{D}_L} \tag{8}$$

being the referential electroelastic tensors. The Einstein convention of summation has been used in Eq. (7), and will be adopted in the following, unless stated otherwise.

The incremental time-harmonic wave motion can then be completely determined from Eqs. (4)-(8) subjected to certain constraints imposed on the boundary of the electroelastic body in the initial configuration, which will be addressed in Section 5.

### 3. Nonlinear Deformation of a Homogeneous DE Cylinder

Consider the homogeneous compressible DE solid cylinder in Fig. 1 with its radius and length in the undeformed configuration being $R_0$ and $L$, respectively. An electric voltage difference $V$ is applied to the mechanically negligible electrodes at the top and bottom surfaces of the cylinder, which is simultaneously subjected to an axial force $N$. As a result, the radius and the length of the cylinder become $r_0$ and $l$ in the deformed configuration, respectively.

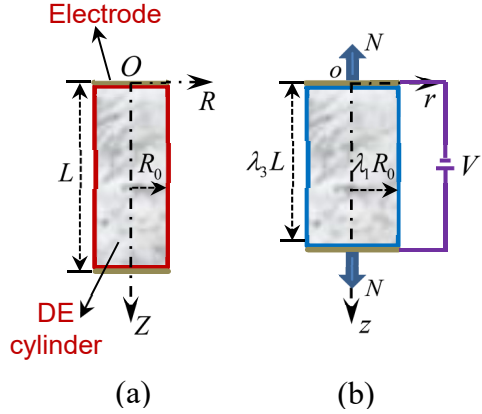

**Fig. 1.** Diagram of a soft DE cylinder with mechanically negligible electrodes bonded onto the two end (top and bottom) surfaces: (a) undeformed configuration, along with the reference coordinates $(R,\Theta,Z)$; (b) deformed configuration induced by an axial electric voltage $V$ and an axial force $N$, along with the current coordinates $(r,\theta,z)$.

This process involves a uniform nonlinear deformation, which can be described in two cylindrical systems $(R,\Theta,Z)$ and $(r,\theta,z)$ by

$$r = \lambda_1 R, \quad \theta = \Theta, \quad z = \lambda_3 Z \qquad (9)$$

where $\lambda_1$ and $\lambda_3$ are the radial and axial principal stretches, respectively. Thus, the deformation gradient tensor can be expressed as $\mathbf{F} = \mathrm{diag}[\lambda_1,\lambda_1,\lambda_3]$. The only non-zero component of the Eulerian electric displacement vector $\mathbf{D}$ and that of the electric field vector $\mathbf{E}$ are the axial components $D_3$ and $E_3$. Accordingly, we have $D_{L3} = J\lambda_3^{-1}D_3$ and $E_{L3} = \lambda_3 E_3$. Using $\mathbf{S} = J\mathbf{F}^{-1}\boldsymbol{\sigma}$, the constitutive relations (1) can be expressed in terms of the three principal stretches and the axial nominal electric displacement component as [46-48]

$$\sigma_k = J^{-1}\lambda_k \frac{\partial W(\lambda_1,\lambda_2,\lambda_3,D_{L3})}{\partial \lambda_k}, \quad E_{L3} = \frac{\partial W(\lambda_1,\lambda_2,\lambda_3,D_{L3})}{\partial D_{L3}}, \quad \text{(no summation over } k\text{)} \qquad (10)$$

where $J = \lambda_1\lambda_2\lambda_3$, and $\sigma_k (k=1,2,3)$ are the principal Cauchy stress components.

Most of the existing researches on soft DE PCs take no account of material compressibility [29-34]. In order to analyze the longitudinal wave propagation, the compressibility must be considered [8, 49]. Here, we adopt the following ideal compressible DE model [49]:

$$W = W_{elas}(\lambda_1, \lambda_2, \lambda_3) + \frac{1}{2\varepsilon J}\lambda_3^2 D_{L3}^2 \tag{11}$$

where $\varepsilon = \varepsilon_0 \varepsilon_r$ is the material permittivity with $\varepsilon_0 = 8.85$ pF/m and $\varepsilon_r$ being the relative permittivity, and the part $W_{elas}(\lambda_1, \lambda_2, \lambda_3)$ is purely elastic. To describe the behavior of the compressibility, $W_{elas}(\lambda_1, \lambda_2, \lambda_3)$ is chosen to be of the following compressible Gent model [23, 50]:

$$W_{elas}^G = -\frac{\mu J_m}{2}\ln\left(1 - \frac{I_1 - 3}{J_m}\right) - \mu \ln J + \left(\frac{\Lambda}{2} - \frac{\mu}{J_m}\right)(J-1)^2 \tag{12}$$

where $I_1 = \lambda_1^2 + \lambda_2^2 + \lambda_3^2$, $\mu$ and $\Lambda$ are the shear modulus and the first Lamé's parameter in the undeformed configuration. The bulk modulus is then calculated as $K = \Lambda + 2\mu/3$. The Gent model can account for the particular strain-stiffening effect. The parameter $J_m$ in Eq. (12) is the dimensionless Gent constant reflecting the limiting chain extensibility of rubber networks [51] such that the elastic strain energy approaches infinity in the limit $I_1 - 3 \to J_m$. In addition, the Gent model (12) can reduce to the neo-Hookean model in the limit $J_m \to \infty$. In the following, we will give, just for illustration, the analytical formulae for the Gent model. The analytical results for the neo-Hookean model, which can be obtained from those for the Gent model by setting $J_m \to \infty$, are explicitly given in the appendix for the convenience of reference.

Substituting Eqs. (11) and (12) into Eq. (10) and using the relations $D_{L3} = J\lambda_3^{-1}D_3$ and $E_{L3} = \lambda_3 E_3$, we obtain the principal Cauchy stress components and the axial Eulerian electric field component as

$$\begin{aligned}\sigma_1 &= \sigma_2 = \mu J^{-1}\left(\frac{J_m}{J_m - I_1 + 3}\lambda_1^2 - 1\right) + \left(\Lambda - \frac{2\mu}{J_m}\right)(J-1) - \frac{D_3^2}{2\varepsilon}, \\ \sigma_3 &= \mu J^{-1}\left(\frac{J_m}{J_m - I_1 + 3}\lambda_3^2 - 1\right) + \left(\Lambda - \frac{2\mu}{J_m}\right)(J-1) + \frac{D_3^2}{2\varepsilon}, \quad E_3 = \frac{D_3}{\varepsilon}\end{aligned} \tag{13}$$

Because of the curl-free electric field (i.e., $\text{curl}\,\mathbf{E} = \mathbf{0}$), we have the relation $\mathbf{E} = -\text{grad}\phi$ with $\phi$ being the electrostatic potential. Therefore, using Eq. (13)$_3$, the electric boundary condition imposed on the two ends of the cylinder can be written as

$$V = \phi(l) - \phi(0) = -\int_0^{\lambda_3 L} E_3 dz = -\frac{D_3}{\varepsilon} \lambda_3 L \tag{14}$$

It can be seen that the axial electric displacement is linearly related to the applied electric voltage. Besides, the traction-free boundary condition on the cylindrical surface $r = r_0$ and the condition that the resultant axial force equals $N$ are expressed as

$$\sigma_1 = 0 \tag{15}$$

and

$$\sigma_3 = \frac{N}{\pi r_0^2} = \frac{N}{\pi \lambda_1^2 R_0^2} \tag{16}$$

respectively. Owing to the uniform stress and electric fields in the DE cylinder, the balance laws (2) and (3) are satisfied automatically. Thus, Eqs. (13)-(16) can be used to completely determine the two induced principal stretches $\lambda_1$ and $\lambda_3$ in terms of the axial electric voltage $V$ and the axial force $N$.

## 4. Deformed Configuration of the DE Phononic Cylinder

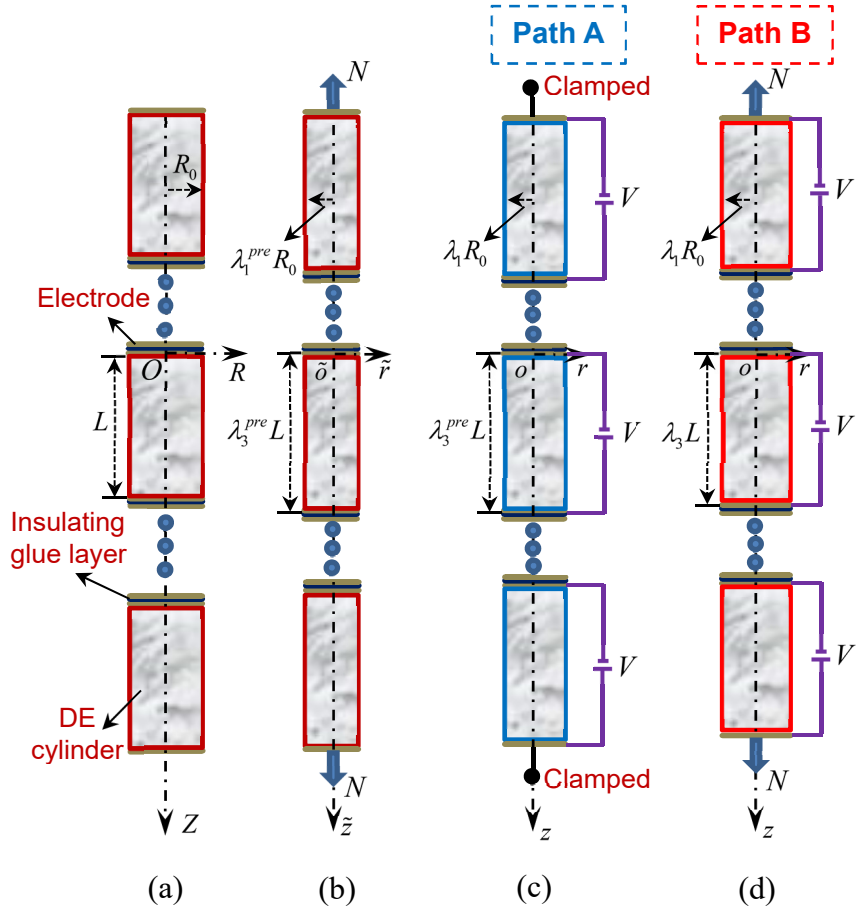

**Fig. 2.** Geometry and actuation of an infinitely long PC made of identical DE sub-cylinders separated by mechanically negligible insulating glue layers: (a) undeformed configuration; (b) pre-stretched configuration induced by an axial force $N$ only; (c) Path A configuration: fixed pre-stretch combined

with a periodically applied electric voltage $V$; (d) Path B configuration: fixed axial force combined with a periodically applied electric voltage $V$.

In the previous section, we have investigated the nonlinear response of a homogeneous DE solid cylinder. Now the interest is turned to consider the nonlinear deformation of the soft DE phononic cylinder with periodic electric boundary conditions under the combined action of the axial electric voltage and the axial force, as indicated in Fig. 2. It is worth mentioning that each identical homogeneous DE sub-cylinder with mechanically negligible electrodes at its top and bottom surfaces represents a unit cell of the soft PC. These sub-cylinders are connected by a small amount of insulating glue (such as epoxy) layer, as shown in Fig. 2.

First, only an axial force $N$ is applied to the undeformed soft DE cylinder (Fig. 2(a)) to form a uniformly pre-stretched configuration (Fig. 2(b)). Utilizing Eqs. (13), (15) and (16) and noticing that the axial electric displacement component is zero, we have

$$\bar{\sigma}_1 = \frac{1}{J^{pre}}\left[\frac{J_m}{J_m - I_1^{pre} + 3}\left(\lambda_1^{pre}\right)^2 - 1\right] + \left(\bar{\Lambda} - \frac{2}{J_m}\right)\left(J^{pre} - 1\right) = 0,$$

$$\bar{\sigma}_3 = \frac{1}{J^{pre}}\left[\frac{J_m}{J_m - I_1^{pre} + 3}\left(\lambda_3^{pre}\right)^2 - 1\right] + \left(\bar{\Lambda} - \frac{2}{J_m}\right)\left(J^{pre} - 1\right) = \frac{\bar{N}}{\left(\lambda_1^{pre}\right)^2} \quad (17)$$

where $\bar{\Lambda} = \Lambda/\mu$, $\lambda_1^{pre}$ and $\lambda_3^{pre}$ are the pre-stretches, $\bar{\sigma}_k = \sigma_k/\mu$ and $\bar{N} = N/(\pi\mu R_0^2)$ are the dimensionless principal Cauchy stress components and the dimensionless axial force. Consequently, the pre-stretches $\lambda_1^{pre}$ and $\lambda_3^{pre}$ can be determined from Eq. (17) in terms of $\bar{N}$. The length of the single homogeneous solid cylinder (i.e., the sub-cylinder) becomes $\lambda_3^{pre} L$ in this pre-stretched configuration (Fig. 2(b)).

Based on the pre-stretched configuration (Fig. 2(b)), two different loading paths are considered to form the effective periodicity and to arrive at the desired initial deformed configurations (Figs. 2(c) and 2(d)) of the soft electroactive PC cylinder.

### 4.1 Path A: Fixed Axial Pre-stretch

For loading path A, the axial pre-stretch is kept unchanged, and an identical axial electric voltage is applied to each sub-cylinder to reach the initial deformed configuration (Fig. 2(c)). First, we have the fixed axial pre-stretch condition

$$\lambda_3 = \lambda_3^{pre} \quad (18)$$

Then, inserting Eq. (13)₁ into Eq. (15) and using the relation (14), we can obtain the traction-free

boundary condition on the cylindrical surface as

$$\bar{\sigma}_1 = \frac{1}{J}\left(\frac{J_m}{J_m - I_1 + 3}\lambda_1^2 - 1\right) + \left(\bar{\Lambda} - \frac{2}{J_m}\right)(J-1) - \frac{\bar{V}^2}{2\lambda_3^2} = 0 \qquad (19)$$

where $\bar{V} = \sqrt{\varepsilon/\mu}\, V/L$ is the dimensionless electric voltage. Thus, the initial radial principal stretch $\lambda_1$ can be achieved from Eq. (19) in terms of the voltage $\bar{V}$ and the axial pre-stretch $\lambda_3 = \lambda_3^{pre}$. Furthermore, the axial stress component and the axial force required to maintain the fixed pre-stretch can be calculated from Eqs. (13)$_2$ and (16) as

$$\bar{\sigma}_3 = \frac{1}{J}\left[\frac{J_m}{J_m - I_1 + 3}(\lambda_3)^2 - 1\right] + \left(\bar{\Lambda} - \frac{2}{J_m}\right)(J-1) + \frac{\bar{V}^2}{2(\lambda_3)^2}, \quad \bar{N} = \bar{\sigma}_3 \lambda_1^2 \qquad (20)$$

with $\lambda_3 = \lambda_3^{pre}$ as indicated above.

**4.2 Path B: Fixed Axial Force**

For loading path B, after the axial force $N$ is applied to form the pre-stretched configuration (Fig. 2(b)), we keep it unchanged and then periodically apply the axial electric voltage $V$ to each sub-cylinder to obtain the initial deformed configuration (Fig. 2(d)). With the help of the relation (14), substitution of Eq. (13) into Eqs. (15) and (16) yields

$$\begin{aligned}\bar{\sigma}_1 &= \frac{1}{J}\left(\frac{J_m}{J_m - I_1 + 3}\lambda_1^2 - 1\right) + \left(\bar{\Lambda} - \frac{2}{J_m}\right)(J-1) - \frac{\bar{V}^2}{2\lambda_3^2} = 0, \\ \bar{\sigma}_3 &= \frac{1}{J}\left(\frac{J_m}{J_m - I_1 + 3}\lambda_3^2 - 1\right) + \left(\bar{\Lambda} - \frac{2}{J_m}\right)(J-1) + \frac{\bar{V}^2}{2\lambda_3^2} = \frac{\bar{N}}{\lambda_1^2}\end{aligned} \qquad (21)$$

which determines $\lambda_1$ and $\lambda_3$ in terms of the axial force $\bar{N}$ and the electric voltage $\bar{V}$.

So far, the exact analysis of the large deformation induced by the axial force and the electric voltage has been completed. These two loading paths give birth to two macroscopically different soft PC cylinders so that we can study the effect of the formative paths on the BGs of the longitudinal waves, to be shown below.

**5. Analysis of Incremental Wave Propagation**

Having obtained the initial state of large deformation in the previous section, we will derive analytically the dispersion relation of the incremental longitudinal waves in the initial deformed DE phononic cylinder. We will also carry out the useful limit analysis of the dispersion relation, which can be used to check the numerical results in Section 6.

For the superimposed longitudinal waves, the incremental constitutive equation (6) can be written in component form as

$$\begin{aligned}
\dot{S}_{011} &= \mathcal{A}_{01111} H_{11} + \mathcal{A}_{01122} H_{22} + \mathcal{A}_{01133} H_{33} + \mathcal{B}_{0113} \dot{D}_{L03}, \\
\dot{S}_{022} &= \mathcal{A}_{02211} H_{11} + \mathcal{A}_{02222} H_{22} + \mathcal{A}_{02233} H_{33} + \mathcal{B}_{0223} \dot{D}_{L03}, \\
\dot{S}_{033} &= \mathcal{A}_{03311} H_{11} + \mathcal{A}_{03322} H_{22} + \mathcal{A}_{03333} H_{33} + \mathcal{B}_{0333} \dot{D}_{L03}, \\
\dot{E}_{L03} &= \mathcal{B}_{0113} H_{11} + \mathcal{B}_{0223} H_{22} + \mathcal{B}_{0333} H_{33} + \mathcal{C}_{033} \dot{D}_{L03}
\end{aligned} \quad (22)$$

In the following derivation, the only approximation made is the assumption of one-dimensional stress state which is well-accepted in the classical rod theory. Thus, we have $\dot{S}_{011} = \dot{S}_{022} = 0$ and can express $H_{11}$ and $H_{22}$ in terms of $H_{33}$ and $\dot{D}_{L03}$ from Eq. (22)$_{1,2}$ as

$$\begin{bmatrix} H_{11} \\ H_{22} \end{bmatrix} = -\begin{bmatrix} \mathcal{P}_{11} & \mathcal{P}_{12} \\ \mathcal{P}_{21} & \mathcal{P}_{22} \end{bmatrix} \begin{bmatrix} H_{33} \\ \dot{D}_{L03} \end{bmatrix} \quad (23)$$

where

$$\begin{aligned}
\mathcal{P}_{11} &= \frac{\mathcal{A}_{02222} \mathcal{A}_{01133} - \mathcal{A}_{01122} \mathcal{A}_{02233}}{\mathcal{A}_{01111} \mathcal{A}_{02222} - \mathcal{A}_{01122}^2}, & \mathcal{P}_{12} &= \frac{\mathcal{A}_{02222} \mathcal{B}_{0113} - \mathcal{A}_{01122} \mathcal{B}_{0223}}{\mathcal{A}_{01111} \mathcal{A}_{02222} - \mathcal{A}_{01122}^2}, \\
\mathcal{P}_{21} &= \frac{-\mathcal{A}_{01122} \mathcal{A}_{01133} + \mathcal{A}_{01111} \mathcal{A}_{02233}}{\mathcal{A}_{01111} \mathcal{A}_{02222} - \mathcal{A}_{01122}^2}, & \mathcal{P}_{22} &= \frac{-\mathcal{A}_{01122} \mathcal{B}_{0113} + \mathcal{A}_{01111} \mathcal{B}_{0223}}{\mathcal{A}_{01111} \mathcal{A}_{02222} - \mathcal{A}_{01122}^2}
\end{aligned} \quad (24)$$

Substituting Eq. (23) into Eq. (22)$_{3,4}$, we obtain

$$\dot{S}_{033} = \mathcal{A}_0^e H_{33} + \mathcal{B}_0^e \dot{D}_{L03}, \quad \dot{E}_{L03} = \mathcal{B}_0^e H_{33} + \mathcal{C}_0^e \dot{D}_{L03} \quad (25)$$

where

$$\begin{aligned}
\mathcal{A}_0^e &= \mathcal{A}_{03333} - \mathcal{A}_{01133} \mathcal{P}_{11} - \mathcal{A}_{02233} \mathcal{P}_{21}, \quad \mathcal{C}_0^e = \mathcal{C}_{033} - \mathcal{B}_{0113} \mathcal{P}_{12} - \mathcal{B}_{0223} \mathcal{P}_{22}, \\
\mathcal{B}_0^e &= \mathcal{B}_{0333} - \mathcal{A}_{01133} \mathcal{P}_{12} - \mathcal{A}_{02233} \mathcal{P}_{22} = \mathcal{B}_{0333} - \mathcal{B}_{0113} \mathcal{P}_{11} - \mathcal{B}_{0223} \mathcal{P}_{21}
\end{aligned} \quad (26)$$

Therefore, Eq. (25) is the reduced incremental constitutive equation in Eulerian form with the reduced instantaneous electroelastic moduli $\mathcal{A}_0^e$, $\mathcal{B}_0^e$ and $\mathcal{C}_0^e$.

For the ideal compressible Gent model characterized by Eqs. (11) and (12), the non-zero components of the dimensionless instantaneous electroelastic tensor $\bar{\mathcal{A}}_0 = \mathcal{A}_0 / \mu$, $\bar{\mathcal{B}}_0 = \mathcal{B}_0 / \sqrt{\mu/\varepsilon}$, and $\bar{\mathcal{C}}_0 = \varepsilon \mathcal{C}_0$ can be derived from Eqs. (7) and (8) as

$$\bar{\mathcal{A}}_{01111} = \bar{\mathcal{A}}_{02222} = \frac{1}{J}\left[\frac{J_m \lambda_1^2}{J_m - I_1 + 3}\left(1 + \frac{2\lambda_1^2}{J_m - I_1 + 3}\right) + 1 + J^2\left(\bar{\Lambda} - \frac{2}{J_m}\right) + J\bar{D}_3^2\right],$$

$$\bar{\mathcal{A}}_{01122} = \frac{2\lambda_1^2 J_m}{\lambda_3 (J_m - I_1 + 3)^2} + \left(\bar{\Lambda} - \frac{2}{J_m}\right)(2J - 1) + \frac{\bar{D}_3^2}{2},$$

$$\bar{\mathcal{A}}_{01133} = \bar{\mathcal{A}}_{02233} = \frac{2\lambda_3 J_m}{(J_m - I_1 + 3)^2} + \left(\bar{\Lambda} - \frac{2}{J_m}\right)(2J - 1) - \frac{\bar{D}_3^2}{2}, \qquad (27)$$

$$\bar{\mathcal{A}}_{03333} = \frac{1}{J}\left[\frac{J_m \lambda_3^2}{J_m - I_1 + 3}\left(1 + \frac{2\lambda_3^2}{J_m - I_1 + 3}\right) + 1 + J^2\left(\bar{\Lambda} - \frac{2}{J_m}\right)\right],$$

and

$$\bar{\mathcal{B}}_{0113} = \bar{\mathcal{B}}_{0223} = -\bar{\mathcal{B}}_{0333} = -\bar{D}_3, \quad \bar{\mathcal{C}}_{033} = 1 \qquad (28)$$

where $\bar{D}_3 = D_3 / \sqrt{\mu\varepsilon}$ is the dimensionless electric displacement, which is related to $\bar{V}$ as $\bar{D}_3 = \bar{V}/\lambda_3$ according to Eq. (14).

Here, using $w$ to denote the incremental axial displacement, we have $H_{33} = dw/dz$. Thus, the incremental governing equations (4) and (5)$_2$ in Eulerian form become

$$\dot{S}_{033,3} = \rho w_{,tt}, \quad \dot{D}_{L03,3} = 0 \qquad (29)$$

Note that due to the assumption of one-dimensional stress state and the applied axial electric voltage, the increments $\dot{S}_{011}$, $\dot{S}_{022}$, $\dot{D}_{L01}$ and $\dot{D}_{L02}$ are assumed to vanish. In addition, the lateral inertial effect is neglected. The inclusion of lateral inertial effect will lead to the Love rod theory, which exceeds the scope of the present study.

According to Eq. (29)$_2$, the incremental axial electric displacement is uniform along each DE sub-cylinder, i.e., $\dot{D}_{L03} = \text{constant}$ in each sub-cylinder. Inserting Eq. (25)$_1$ into Eq. (29)$_1$ leads to the incremental wave equation

$$\frac{d^2 w}{dz'^2} + \varpi^2 \kappa^2 w = 0 \qquad (30)$$

where the harmonic time-dependency $e^{i\omega t}$ with $\omega$ being the angular frequency is assumed for all fields and is omitted in the following, and the dimensionless quantities are defined as

$$\kappa^2 = \frac{\bar{\rho}}{\bar{\mathcal{A}}_0^e}, \quad \bar{\mathcal{A}}_0^e = \frac{\mathcal{A}_0^e}{\mu}, \quad \bar{\rho} = \frac{\rho}{\rho_r}, \quad \varpi^2 = \lambda_3^2 \Omega^2, \quad \Omega^2 = \frac{\rho_r \omega^2}{\mu} L^2 \qquad (31)$$

Note that the axial coordinate $z$ has been scaled by the deformed unit cell length $l = \lambda_3 L$ to form a new, dimensionless axial coordinate $z' = z/l$ in Eq. (30). Thus, we obtain the incremental

displacement in each sub-cylinder as

$$w(z') = M_+ e^{i\varpi\kappa z'} + M_- e^{-i\varpi\kappa z'} \tag{32}$$

where $\varpi\kappa$ is the axial wave number inside the sub-cylinder, and the undetermined complex coefficients $M_+$ and $M_-$ are determined from the incremental boundary conditions. Inserting Eq. (32) in Eq. (25)$_1$ gives the axial incremental stress as

$$\dot{S}_{033}(z') = i\varpi\kappa \mathcal{A}_0^e \left( M_+ e^{i\varpi\kappa z'} - M_- e^{-i\varpi\kappa z'} \right)/l + \mathcal{B}_0^e \dot{D}_{L03} \tag{33}$$

Owing to the electric voltage periodically applied to each sub-cylinder (see Figs. 2(c) and 2(d)), the formed structure is periodic along the axial direction and becomes a DE phononic cylinder with the unit cell length $l = \lambda_3 L$ for both two loading paths. Using the Bloch-Floquet theorem, we can express the relation for the incremental displacement $w$ and stress $\dot{S}_{033}$ at the interfaces delimiting the unit cell ($z' = 0$ and $z' = 1$) as

$$w(1) = w(0)s, \quad \dot{S}_{033}(1) = \dot{S}_{033}(0)s \tag{34}$$

where $s = e^{iq}$ with $q$ being the Bloch wave number associated with the longitudinal waves propagating in the DE phononic cylinder. By introducing an incremental electric potential $\dot{\phi}$ such that $\dot{\mathbf{E}}_{L0} = -\text{grad}\,\dot{\phi}$, the incremental Faraday's law (5)$_1$ is satisfied identically. During the incremental motion, the electric voltage applied to two ends of a unit cell remains unchanged, which means that the difference of the incremental electric potential $\dot{\phi}$ between the two ends vanishes, i.e.,

$$\begin{aligned}\dot{V} &= \dot{\phi}(l) - \dot{\phi}(0) = -\int_0^l \dot{E}_{L03} dz = -\int_0^1 \dot{E}_{L03} l dz' \\ &= -\mathcal{B}_0^e \left[ w(1) - w(0) \right] - \mathcal{C}_0^e l \dot{D}_{L03} = 0\end{aligned} \tag{35}$$

where Eq. (25)$_2$ has been used. It should be emphasized that because of the Gauss's law, both the initial and the incremental axial electric displacement are uniform in each DE sub-cylinder. At the interfaces separating the unit cells, the initial electric displacement is continuous, and there is no resultant free electric charge on the two electrodes at the interface; however, the incremental one may be discontinuous from one sub-cylinder to the adjacent one, and there are usually resultant free electric charges on the two electrodes during the wave motion [52]. In fact, the incremental electric displacement in the neighbouring sub-cylinders are related by a factor $s$ from Eqs. (34)$_1$ and (35).

The relation (34) combined with Eqs. (32), (33) and (35) leads to two linear homogeneous algebraic

equations for the undetermined constants $M_+$ and $M_-$ as

$$M_+ \left(e^{i\varpi\kappa} - s\right) + M_- \left(e^{-i\varpi\kappa} - s\right) = 0,$$

$$M_+ \left[\left(e^{i\varpi\kappa} - s\right) - \frac{1-s}{i\varpi\kappa\bar{\mathcal{A}}_0^e}\gamma_{11}\left(e^{i\varpi\kappa} - 1\right)\right] - M_- \left[\left(e^{-i\varpi\kappa} - s\right) + \frac{1-s}{i\varpi\kappa\bar{\mathcal{A}}_0^e}\gamma_{11}\left(e^{-i\varpi\kappa} - 1\right)\right] = 0 \quad (36)$$

where $\gamma_{11} = (\bar{\mathcal{B}}_0^e)^2 / \bar{\mathcal{C}}_0^e$ with $\bar{\mathcal{B}}_0^e = \mathcal{B}_0^e / \sqrt{\mu/\varepsilon}$ and $\bar{\mathcal{C}}_0^e = \varepsilon\mathcal{C}_0^e$. As a necessary condition for nontrivial solutions, the determinant of the coefficient matrix of Eq. (36) should vanish, which yields the following quadratic characteristic equation:

$$s^2\left[1 - \alpha\frac{\sin(\kappa\varpi)}{\kappa\varpi}\right] - 2s\left[\cos(\kappa\varpi) - \alpha\frac{\sin(\kappa\varpi)}{\kappa\varpi}\right] + 1 - \alpha\frac{\sin(\kappa\varpi)}{\kappa\varpi} = 0 \quad (37)$$

where $\alpha = \gamma_{11} / \bar{\mathcal{A}}_0^e$. Note that Eq. (37) has either real or complex conjugate solutions. Real solutions of $s$ correspond to stop band and in this case, $q$ is a complex wave number with a real part equal to 0 or $\pi$. In contrast, complex conjugate solutions correspond to pass band with real wave number $q$ [52]. Supposing $1 - \alpha\tan(\kappa\varpi/2)/(\kappa\varpi/2) > 0$, we can write the complex conjugate solutions as

$$s = \left[\cos(\kappa\varpi) - \alpha\frac{\sin(\kappa\varpi)}{\kappa\varpi}\right] \Big/ \left[1 - \alpha\frac{\sin(\kappa\varpi)}{\kappa\varpi}\right]$$

$$\pm i\left[\sin(\kappa\varpi)\sqrt{1 - \alpha\frac{\tan(\kappa\varpi/2)}{\kappa\varpi/2}}\right] \Big/ \left[1 - \alpha\frac{\sin(\kappa\varpi)}{\kappa\varpi}\right] \quad (38)$$

The real part of Eq. (38) gives the dispersion relation of the pass band as

$$\cos q = \left[\cos(\kappa\varpi) - \alpha\frac{\sin(\kappa\varpi)}{\kappa\varpi}\right] \Big/ \left[1 - \alpha\frac{\sin(\kappa\varpi)}{\kappa\varpi}\right] \quad (39)$$

The imaginary part of Eq. (38) is then satisfied automatically.

The parameters $\alpha$ and $\kappa\varpi$ in Eq. (39) are related to the reduced instantaneous electroelastic moduli $\mathcal{A}_0^e$, $\mathcal{B}_0^e$ and $\mathcal{C}_0^e$, which are determined by the stretches $\lambda_1$, $\lambda_3$ and the electric displacement component $\bar{D}_3$ from Eqs. (24) and (26)-(28). Consequently, by adjusting the applied axial force and the electric voltage, we can change the instantaneous electromechanical properties of the DE phononic cylinder, which in turn has a paramount effect on the longitudinal wave propagation characteristics (e.g., the band structure).

When there is no electric voltage applied to the soft PC (i.e., $\bar{V} = 0$), we have $\alpha = 0$, which gives $\cos q = \cos(\kappa\varpi)$ from Eq. (39). Thus, we have the solution $q = \pm\kappa\varpi$ for which the wave is non-dispersive even when there exists a pure pre-stretch by the axial force only (seeing Fig. 2(b)). That is to say, purely propagative solutions are obtained for the incremental displacement $w$ and no reflection takes place at the electrodes of the interfaces [52].

Furthermore, the frequency limits of the BG at the border of the first Brillouin zone $q = \pm \pi$ can be determined from Eq. (39), which now reads

$$\left[\cos(\kappa\varpi) - \alpha \frac{\sin(\kappa\varpi)}{\kappa\varpi}\right] \Big/ \left[1 - \alpha \frac{\sin(\kappa\varpi)}{\kappa\varpi}\right] = -1 \tag{40}$$

This equation leads to $\cos(\kappa\varpi/2) = 0$ or $1 - \alpha \tan(\kappa\varpi/2)/(\kappa\varpi/2) = 0$, which in turn give rise to

$$\begin{aligned} \kappa\varpi &= (2m+1)\pi \quad (m = 0,1,2,\cdots), \\ \tan(\kappa\varpi/2) &= \kappa\varpi/(2\alpha) \end{aligned} \tag{41}$$

In a similar manner, the frequency limits at the center of the first Brillouin zone $q = 0$ can be obtained by setting $\cos q = 1$ in Eq. (39), which gives

$$\kappa\varpi = 2m\pi \quad (m = 0,1,2,\cdots) \tag{42}$$

In the long wavelength limit, the effective wave velocity of the longitudinal waves can be calculated by utilizing the second order Taylor expansion of Eq. (39) as

$$V_{eff}^2 = \omega^2/(q/l)^2 = (1-\alpha)J\bar{\mathcal{A}}_0^e c_T^2 \tag{43}$$

where $c_T = \sqrt{\mu/\rho_r}$ is the shear wave velocity in the undeformed state. Furthermore, if there is no pre-stretch and electric voltage applied to the soft PC, we have $\alpha = 0$, $J = 1$, and $\bar{\mathcal{A}}_0^e = (2+3\bar{\Lambda})/(1+\bar{\Lambda})$. The long wavelength limit (43) becomes

$$V_{eff}^2 = \frac{2\mu + 3\Lambda}{\mu + \Lambda} c_T^2 = \frac{E}{\rho_r} \tag{44}$$

where $E = \mu(2\mu+3\Lambda)/(\mu+\Lambda)$ is the Young's modulus of the undeformed soft PC cylinder. The result (44) is in accordance with the classical rod theory [53].

It is noted that if the applied electric voltage is beyond certain critical value $\bar{V}_c$ such that $\alpha > 1$, the argument $\kappa\varpi/2$ appearing in the frequency limit (41)$_2$ has no solution in the range of $(0, \pi/2)$ and the effective wave velocity in the long wavelength limit (43) takes a purely imaginary value. The underlying mechanism should be that the soft dielectric PCs are susceptible to various kinds of failure [54] (such as buckling, electromechanical instability (EMI) and electric breakdown (EB)), imposing corresponding limits on the external stimuli. This becomes particularly clear in the case of Path B as to be shown below. Therefore, a stability condition $\alpha < 1$ should be satisfied and the applied electric voltage generally should be less than the critical voltage $\bar{V}_c$ determined from the condition $\alpha = 1$.

## 6. Numerical Results and Discussions

We are now interested in knowing how the applied electric voltage and the axial force affect the incremental longitudinal wave propagation in the deformed DE phononic cylinder. The commercial product Fluorosilicone 730 [31] is chosen as the DE, with its material properties given by $\rho_0 = 1400 \text{ kg/m}^3$, $\mu = 167.67 \text{ kPa}$, $\varepsilon_r = 7.11$, and $E_{EB} = 372 \text{ MV/m}$, where $E_{EB}$ is the dielectric strength beyond which the phenomenon of EB occurs. In addition, the first Lamé's parameter $\Lambda$ and the Gent constant $J_m$ are consistently set to be $\Lambda = 100\mu$ and $J_m = 10$, respectively, in the following calculations. The radius and length of the homogeneous DE cylinder in the undeformed configuration are $R_0 = 10 \text{ mm}$ and $L = 50 \text{ mm}$, respectively. As mentioned earlier, the results of the neo-Hookean model can be obtained by simply taking $J_m \to \infty$.

During the loading process from Fig. 2(a) to Fig. 2(b) under the axial force $N$ only, we plot in Fig. 3 the pre-stretches $\lambda_1^{pre}$ and $\lambda_3^{pre}$ based on Eqs. (A3) and (17) as functions of the dimensionless resultant axial force $\bar{N}$ for both neo-Hookean and Gent models. It is found that the predictions of the two nonlinear material models are almost indistinguishable when $\bar{N} \leq 1$. However, for $\bar{N} > 1$, the difference between them gradually increases. Specifically, owing to the existence of strain-stiffening effect, the axial pre-stretch $\lambda_3^{pre}$ based on the Gent model is smaller than that predicted by the neo-Hookean model under the same axial force.

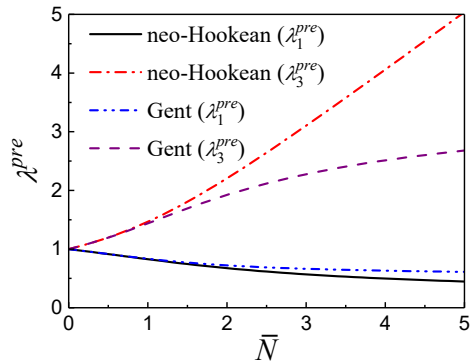

**Fig. 3.** Variations of pre-stretches $\lambda_1^{pre}$ and $\lambda_3^{pre}$ as functions of the dimensionless resultant axial force $\bar{N}$ in the DE phononic cylinder in the absence of the electric voltage for both neo-Hookean and Gent

models.

## 6.1 Path A: Fixed Axial Pre-stretch

For the uniform axisymmetric deformation of the DE cylinder, the voltage corresponding to EB is given by $V_{EB} = E_{EB}l = E_{EB}\lambda_3 L$ which results in the dimensionless EB voltage as $\bar{V}_{EB} = E_{EB}\lambda_3\sqrt{\varepsilon/\mu}$. As previously mentioned, there exists a critical electric voltage $\bar{V}_c$ corresponding to $\alpha = 1$ which makes the frequency limit (41)$_2$ and the effective wave velocity (43) vanish. Note that for Path A, the axial pre-stretch is the same as the initial axial stretch, and hence there is no need to have the superscript 'pre' on $\lambda_3$ in the following discussions.

For Path A with fixed axial pre-stretch $\lambda_3$, the variations of the two voltages $\bar{V}_c$ and $\bar{V}_{EB}$ with the axial pre-stretch in the DE phononic cylinder are displayed in Fig. 4 for both neo-Hookean and Gent models. Since the axial pre-stretch is kept unchanged, the two material models provide the same EB voltage. It can be seen that the critical voltages given by the two models are far below the EB voltage, and hence the allowable range of the voltage is assumed to be $(0, \bar{V}_c)$ for loading path A. Besides, the critical voltages given by the two models increase with the axial pre-stretch, and the critical voltage for the Gent model is remarkably larger than that for the neo-Hookean model especially when $\lambda_3 > 2$.

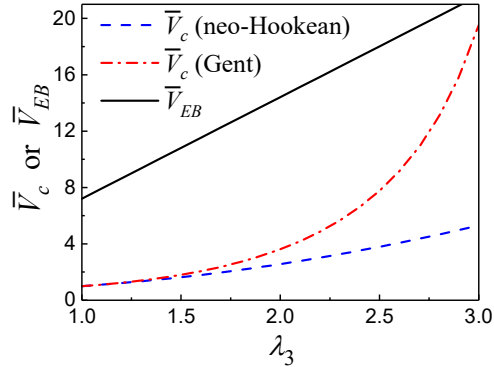

**Fig. 4.** Dimensionless critical voltages $\bar{V}_c$ corresponding to $\alpha = 1$ and dimensionless EB voltages $\bar{V}_{EB}$ versus the axial pre-stretch $\lambda_3$ in the DE phononic cylinder (Path A) for both neo-Hookean and Gent models.

In Fig. 5, based on Eqs. (A4) and (19), the nonlinear response of the initial radial stretch $\lambda_1$, the axial stress $\bar{\sigma}_{33}$, and the axial force $\bar{N}$ in the DE phononic cylinder to the dimensionless electric voltage $\bar{V}$ is presented for both neo-Hookean and Gent models at three different axial pre-stretches (i.e., $\lambda_3 = 1$, 1.5, and 3) for loading path A. Note that, the critical voltage shown in Fig. 5 varies from case to case, depending on the axial pre-stretch and the material model. As we can see from Figs. 5(a) and 5(d) that the radial stretch is kept almost unchanged with the applied voltage except for $\lambda_3 = 3$ for the Gent model, where the radial stretch increases a little with the voltage. Therefore, the DE cylinder can be envisioned as nearly incompressible with the Poisson's ratio $\nu$ approximately equal to 0.495 based on the relations $\Lambda = 100\mu$ and $\nu = \Lambda/[2\Lambda + \mu]$. Additionally, the axial stress and the axial force for both material models increase monotonically with the voltage, which is reasonable because when the voltage is applied to the DE cylinder it intends to shorten itself. In the allowable voltage range, the axial stress, the axial force, and the amount of their increase for the Gent model are larger than those for the neo-Hookean model, especially for $\lambda_3 = 3$, due to the strain-stiffening effect of the Gent model.

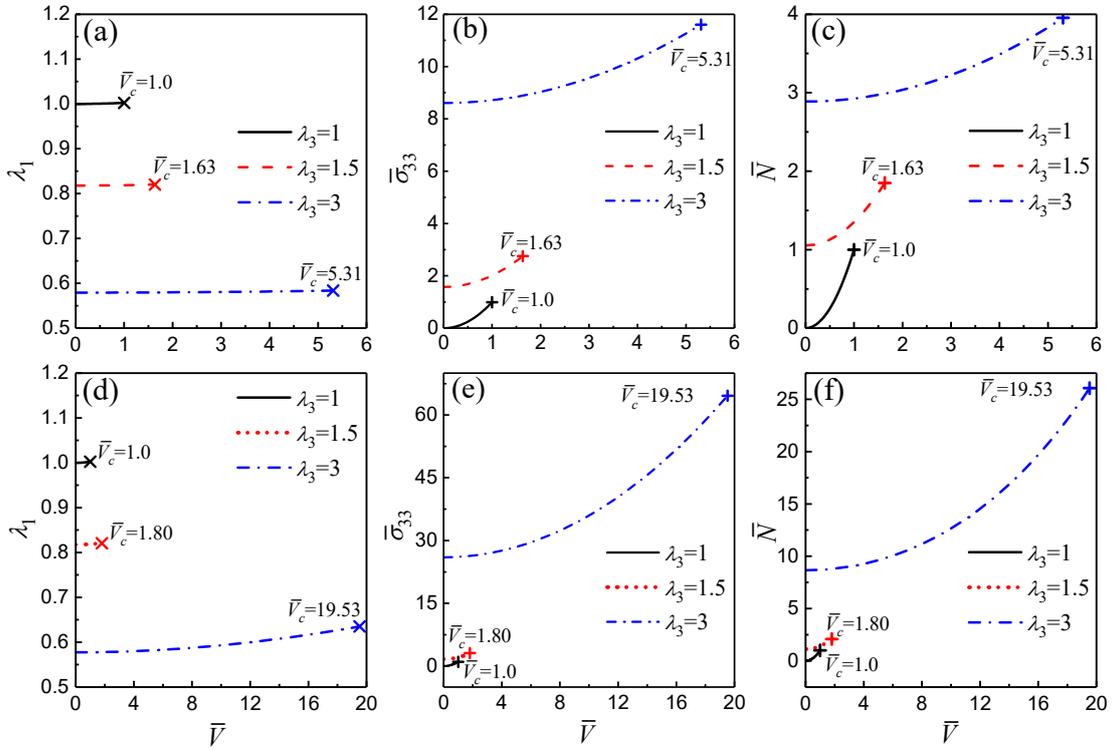

**Fig. 5.** Variations of the initial radial stretch $\lambda_1$ (a, d), the axial stress $\bar{\sigma}_{33}$ (b, e), and the axial force $\bar{N}$ (c, f) as functions of the dimensionless electric voltage $\bar{V}$ in the DE phononic cylinder (Path A) at different axial pre-stretches for both neo-Hookean (a-c) and Gent (d-f) models.

For three different pre-stretches $\lambda_3 = 1$, 1.5, and 3 of loading path A, the dispersion diagrams of the incremental longitudinal waves in the DE phononic cylinder are depicted in Figs. 6(a)-6(c), respectively, for the neo-Hookean model at different electric voltages lower than the critical value. It is observed from Fig. 6 that, in the absence of an electric voltage ($\bar{V} = 0$), there is no BG and the wave propagation is non-dispersive even when a pure pre-stretch is induced in the DE cylinder by the axial force only, as pointed out previously. When the voltage is applied to the DE phononic cylinder, the band structure exhibits BGs at the border of the first Brillouin zone ($q = \pi$). The width of BGs gets enlarged when the applied voltage increases, especially when the voltage approaches the critical value where the lower frequency limit of the first BG vanishes. The width of the first BG is obviously larger than that of the second. Consequently, the DE phononic cylinder can be considered as an electrical switching device between non-dispersive waves ($\bar{V} = 0$) and dispersive waves with Bragg band gaps ($\bar{V} \neq 0$). In addition, we notice that no BG appears at the center of the first Brillouin zone ($q = 0$), as evidenced by Eq. (42). The results for the Gent model are qualitatively similar to those for the neo-Hookean model and hence omitted here for simplicity. It shall be noted that the above phenomena are pretty similar to those observed for a hard piezoelectric phononic rod with periodic electrical boundary conditions [52].

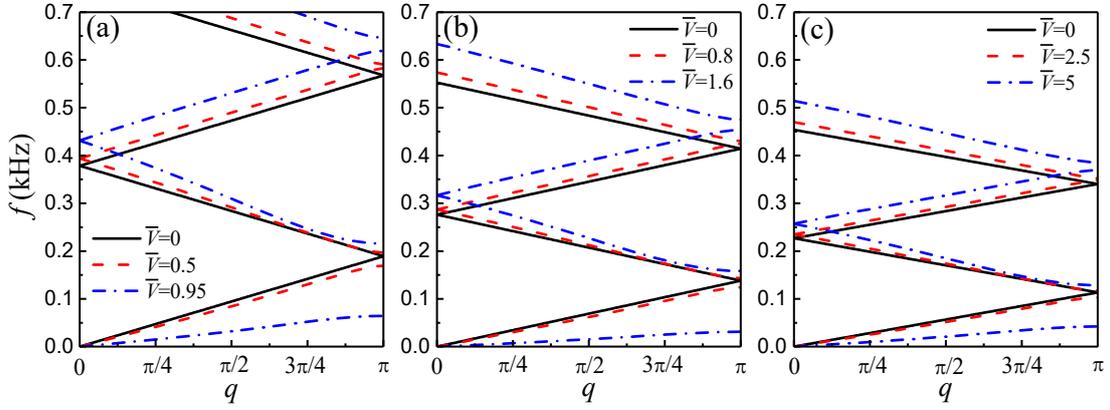

**Fig. 6.** Dispersion diagrams of the incremental longitudinal waves in the DE phononic cylinder (Path A) for the neo-Hookean model at different electric voltages and three different axial pre-stretches: (a) $\lambda_3 = 1$; (b) $\lambda_3 = 1.5$; (c) $\lambda_3 = 3$.

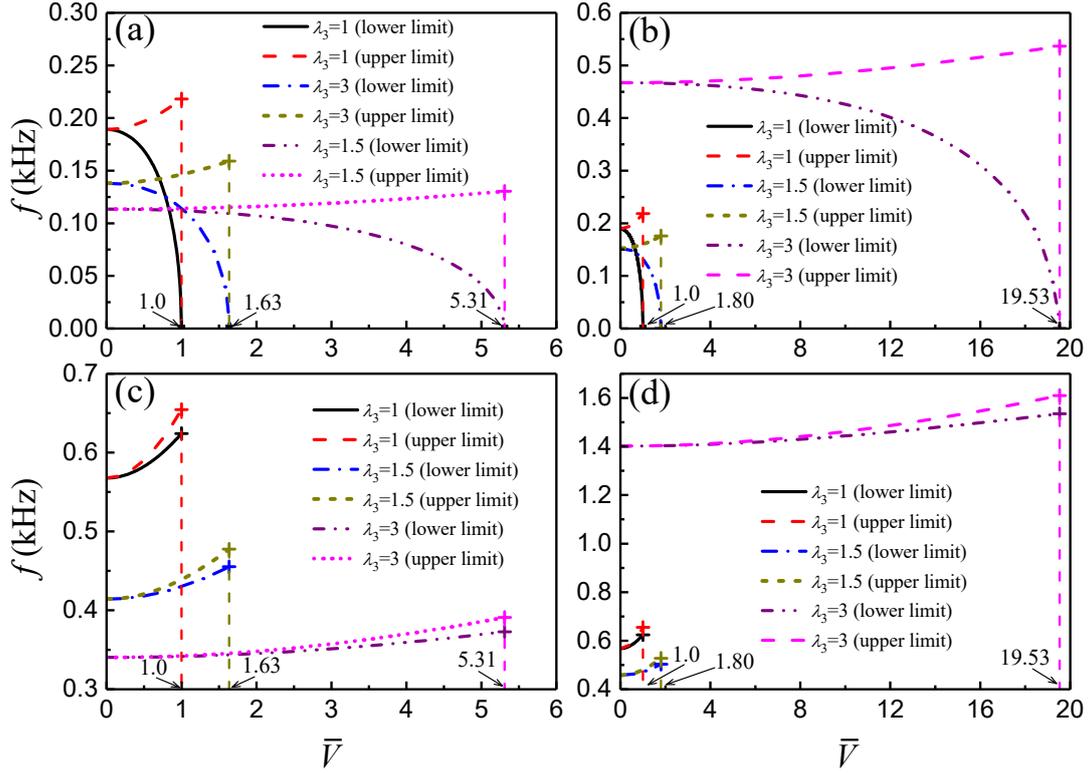

**Fig. 7.** Frequency limits of the first (a, b) and second (c, d) Bragg BGs versus the dimensionless electric voltage $\bar{V}$ in the DE phononic cylinder (Path A) at different axial pre-stretches for both neo-Hookean (a, c) and Gent (b, d) models.

In order to clearly show how the frequency limits of the Bragg BGs vary with the applied voltage, the variations with the voltage of the lower and upper frequency limits of the first two Bragg BGs of the DE phononic cylinder are illustrated in Fig. 7 for both neo-Hookean and Gent models at three different axial pre-stretches $\lambda_3 = 1$, 1.5, and 3 of loading path A. It can be noted from Figs. 7(a) and 7(b) that, for the first BG and the two material models, as the applied voltage increases from zero to the critical value, the upper frequency limit of the band gap continuously goes up while the lower one decreases substantially to zero. As a consequence, the width of the first BG increases considerably with the applied voltage. Specifically, for the neo-Hookean model, the width of the first BG for three different pre-stretches 1, 1.5 and 3 varies from 0kHz to 0.218kHz, 0.159kHz and 0.130kHz, respectively. For the Gent model, the variation ranges of the first BG width for pre-stretches 1, 1.5 and 3 are (0,0.218)kHz, (0,0.176)kHz and (0,0.537)kHz, respectively. Therefore, with the increase of the pre-stretch, the maximal BG width goes down for the neo-Hookean model, while that for the Gent model first decreases and then increases reversely. Additionally, in order to reach the same level of the BG width, a higher electric voltage is required for the increasing axial pre-stretch. For the second BG, it is found from Figs. 7(c) and 7(d) that, its upper and lower frequency limits as well as its width all increase with the voltage for the two material models. Nonetheless, the maximal width of the second BG is significantly smaller than the first one, which is essentially the same as that shown in Fig. 6.

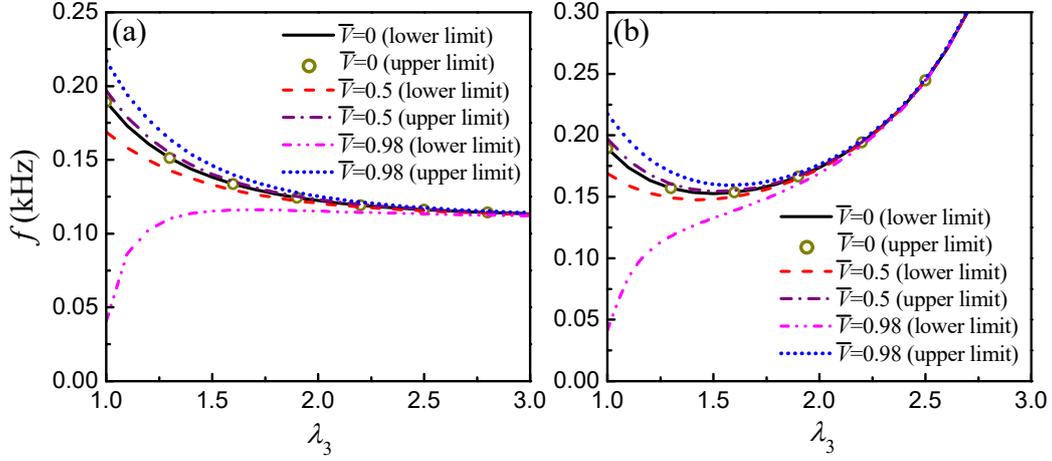

**Fig. 8.** Frequency limits of the first Bragg BG versus the axial pre-stretch $\lambda_3$ in the DE phononic cylinder (Path A) at different electric voltages for both neo-Hookean (a) and Gent (b) models.

Fig. 8 displays the variations with the axial pre-stretch of the lower and upper frequency limits of the first Bragg BG of the DE phononic cylinder for both neo-Hookean and Gent models at three different voltages $\bar{V} = 0$, 0.5, and 0.98 for loading path A. One notes that, in the absence of voltage ($\bar{V} = 0$), the upper and lower frequency limits coincide, i.e., there is no BG. It is seen from Fig. 8(a) that, for the neo-Hookean model, the upper and lower frequency limits as well as the width of the band gap all decrease monotonically with the axial pre-stretch except the lower frequency limit at $\bar{V} = 0.98$. Since the voltage ($\bar{V} = 0.98$) tends to the critical value corresponding to the axial pre-stretch $\lambda_3 = 1$, the lower frequency limit at $\bar{V} = 0.98$ first increases and then declines slowly with the axial pre-stretch. In contrast, Fig. 8(b) shows that, for the Gent model, the frequency limits decrease to a minimum and then increase considerably except the lower frequency limit at $\bar{V} = 0.98$, which increases monotonically with the axial pre-stretch. These results are in accordance with those shown in Figs. 7(a) and 7(b).

The main characteristics of the first tunable Bragg BG for the DE phononic cylinder are concluded in Table 1 for both neo-Hookean and Gent models at three different axial pre-stretches of loading path A. The first BG central frequency in the absence of electric voltage has been utilized for normalization. As we can see from Table 1, with the increase of the axial pre-stretch, the first BG central frequency goes down for the neo-Hookean model, while that for the Gent model first reduces and then increases. Therefore, the applied voltage can largely widen the BGs while the axial pre-stretch mainly change the position of the BGs. In addition, the variation ranges of the normalized central frequency and the normalized BG width are almost the same for the three axial pre-stretches and the two material models, with their respective ranges being (1-0.58) and (0-1.15).

Table 1. Main characteristics of the first tunable Bragg BG for the piezoelectric phononic rod [52] and the DE phononic cylinder at three different axial pre-stretches of both neo-Hookean (N-H) and Gent models (Path A). The first BG central frequency without the applied electrical variable is used for normalization.

| Pre-stretch | Control range of normalized electric voltage ($\bar{V}$) | 1st band gap central frequency (kHz) | Variation of 1st band gap normalized central frequency | 1st band gap width (kHz) | Variation of 1st band gap normalized frequency width |
|---|---|---|---|---|---|
| $\lambda_3 = 1$ | N-H: 0-1.00 | 0.189-0.109 | 1-0.577 | 0-0.218 | 0-1.153 |
|  | Gent: 0-1.00 | 0.189-0.109 | 1-0.577 | 0-0.218 | 0-1.153 |
| $\lambda_3 = 1.5$ | N-H: 0-1.63 | 0.138-0.080 | 1-0.580 | 0-0.159 | 0-1.152 |
|  | Gent: 0-1.80 | 0.153-0.088 | 1-0.575 | 0-0.176 | 0-1.150 |
| $\lambda_3 = 3$ | N-H: 0-5.31 | 0.113-0.065 | 1-0.575 | 0-0.130 | 0-1.150 |
|  | Gent: 0-19.53 | 0.467-0.268 | 1-0.575 | 0-0.537 | 0-1.150 |
| PZT rod | 0-+$\infty$ nF* | 192-160 | 1-0.833 | 0-64 | 0-0.333 |

*Control range of the external capacitance in Ref. [52].

By tuning the external electrical capacitance, Degraeve et al. [52] obtained the main characteristics of the first tunable Bragg BG for a hard piezoelectric phononic crystal rod. Their results are given in the last row of Table 1 for comparison. It is found that, the variation ranges of the normalized central frequency and the normalized BG width for the piezoelectric phononic rod are smaller than those for the DE phononic cylinder, revealing that the tunable range in our work is enlarged obviously as compared with that of Degraeve et al. [52]. Besides, compared to the piezoelectric phononic rod, the position of the BGs can be adjusted by the axial pre-stretch for the DE phononic cylinder. Finally, the frequency of the BGs for the DE phononic cylinder is considerably lower than that for the piezoelectric phononic rod. Consequently, the DE phononic cylinder can be a proper candidate of a low frequency filter.

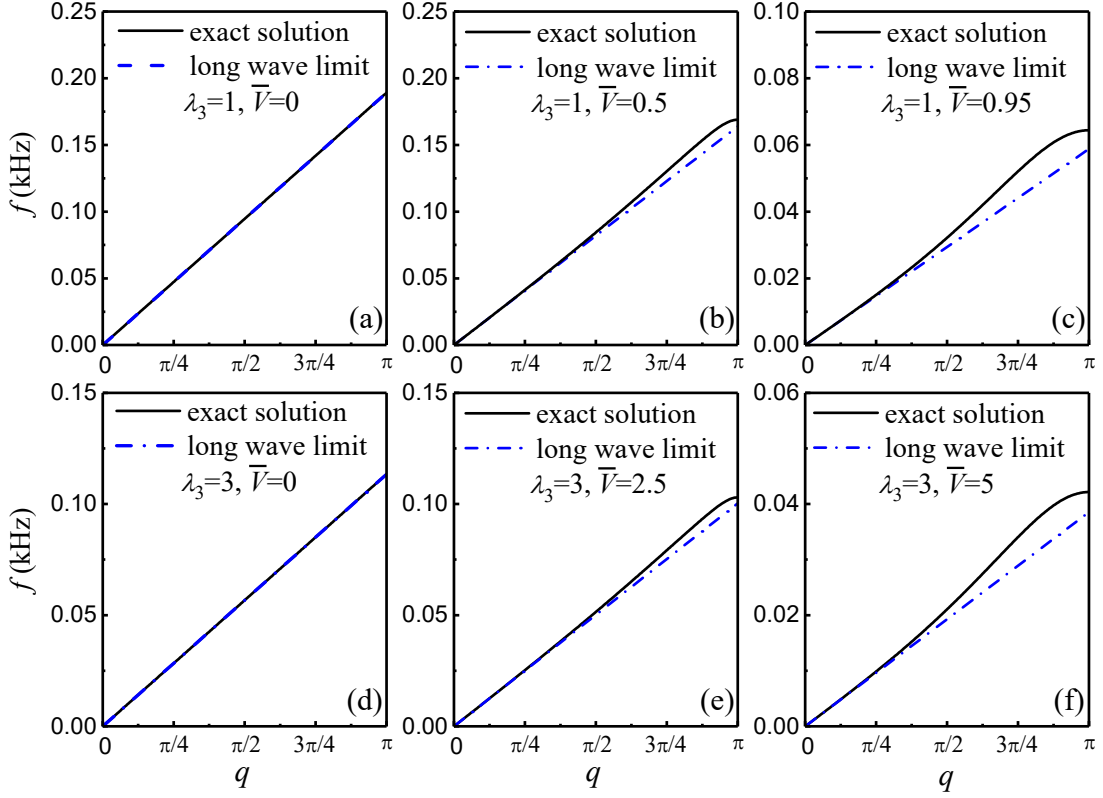

**Fig. 9.** Comparison of the long wavelength limit (43) with the exact solution (39) of the incremental longitudinal waves in the DE phononic cylinder (Path A) at different axial pre-stretches and electric voltages for the neo-Hookean model.

A comparison of the long wave limit (43) with the exact solution (39) is depicted in Fig. 9 for the neo-Hookean model at different axial pre-stretches and electric voltages. It is noted that, in the absence of the electric voltage, the results of the long wave limit and the exact solution are the same in the whole wavenumber range, both predicting the non-dispersive waves. When the voltage is applied, the long wave limits agree well with the exact solution at the long wavelength. However, for a given axial pre-stretch, the accuracy of the long wave limit has a drop with the increasing voltage. For example, at the applied voltages $\bar{V} = 0.5$ and $0.95$ for the axial pre-stretch $\lambda_z = 1$, the long wave limits are in excellent agreement with the exact dispersion curves for $q \leq 5\pi/8$ and $q \leq 3\pi/8$, which correspond to the wavelengths greater than or equal to $3.2l$ and $5.3l$, respectively. The results for the Gent model not shown here are quite analogous.

**6.2 Path B: Fixed Axial Force**

Based on Eq. (A6), the variations of the stretch ratios $\lambda_1/\lambda_1^{pre}$ and $\lambda_3/\lambda_3^{pre}$ with the applied electric voltage $\bar{V}$ are plotted in Fig. 10 for the neo-Hookean model at three different axial forces $\bar{N} =$

0, 2.5 and 5 of loading path B. The pre-stretches $\lambda_1^{pre}$ and $\lambda_3^{pre}$ induced by the axial force only have been used for normalization. It is seen that when the axial stretch decreases and the radial stretch increases, the voltage first increases, reaches a peak value, and then falls. If the voltage is beyond the peak value, there is no axisymmetric deformation solution for the DE cylinder. For loading path B with fixed axial force instead of fixed pre-stretch, the peak value voltage represents the critical state for the onset of EMI [46,55], which is due to the positive feedback between an increasing electric field and a drastically shortening DE cylinder, resulting in the ultimate EB. Furthermore, increasing the axial force enlarges the EMI voltage $\bar{V}_{EMI}$. Specifically, the EMI voltages for axial forces $\bar{N} = 0$, 2.5 and 5 are approximately equal to 0.69, 2.45 and 8.23, respectively.

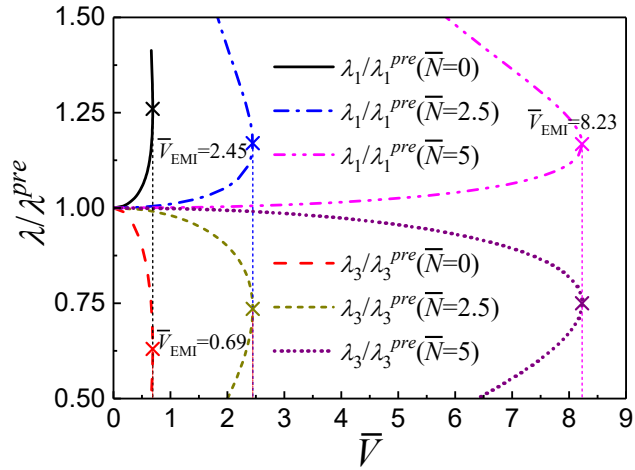

**Fig. 10.** Variations of the radial stretch ratio $\lambda_1/\lambda_1^{pre}$ and the axial stretch ratio $\lambda_3/\lambda_3^{pre}$ as functions of the dimensionless electric voltage $\bar{V}$ in the DE phononic cylinder (Path B) at different axial forces for the neo-Hookean model, where $\lambda_1^{pre}$ and $\lambda_3^{pre}$ are the pre-stretches induced by the axial force only.

For the Gent model, based on Eq. (21), Figs. 11(a)-11(c) illustrate the nonlinear response of the radial stretch in the DE phononic cylinder to the applied electric voltage at three different axial forces $\bar{N} = 0$, 2.5 and 5, respectively, of loading path B. The results for the neo-Hookean model are also shown in Fig. 11 for comparison. It is noted from Fig. 11(a) that, when the cylinder is free from axial force ($\bar{N} = 0$), the nonlinear response predicted by the neo-Hookean model agrees well with that based on the Gent model for $\lambda_1 \leq 1.2$. Similar to the neo-Hookean model, as the radial stretch increases, the voltage rises to a peak value and then falls because a higher electric field is generated by the same voltage when the cylinder shortens. The Gent model gives a slightly higher peak value of the voltage than the neo-Hookean model. This peak value of voltage is also referred to as the EMI voltage $\bar{V}_{EMI}$. However, different from the neo-Hookean model, the voltage given by the Gent model increases again in order to enlarge the radial stretch after falling to a minimum, which is a result of the strain-stiffening effect. Thus, the DE phononic cylinder described by the Gent model may undergo snap-through instability [32,55], which can be potentially exploited to achieve sudden transitions in the BGs. This will be discussed shortly.

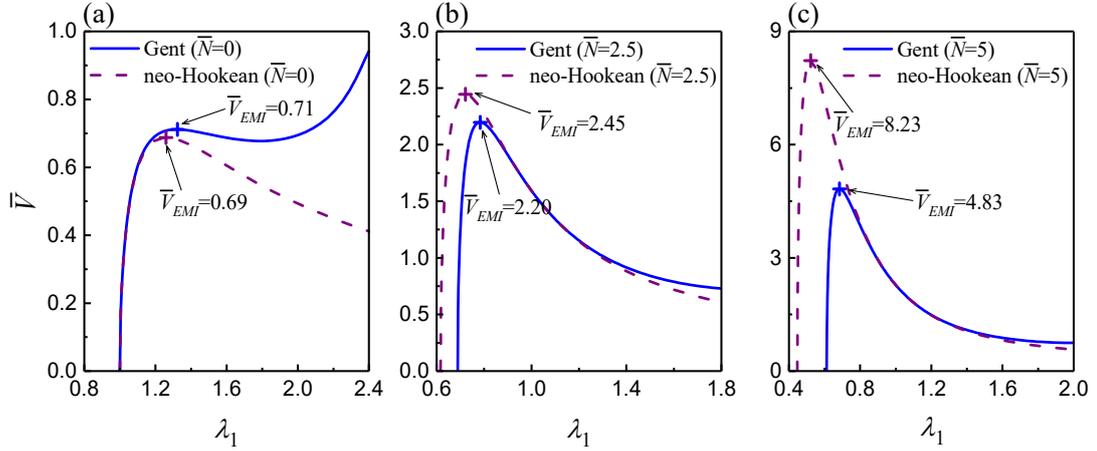

**Fig. 11.** Dimensionless electric voltage $\bar{V}$ versus the radial stretch $\lambda_1$ in the DE phononic cylinder (Path B) for both neo-Hookean and Gent models at three different axial forces: (a) $\bar{N}=0$; (b) $\bar{N}=2.5$; (c) $\bar{N}=5$.

In addition, we can see from Figs. 11(b) and 11(c) that, the tendency of the nonlinear response predicted by the Gent model is pretty similar to that of the neo-Hookean model in the radial stretch range of interest as shown in Figs. 11(b) and 11(c) for $\bar{N}=2.5$ and 5. However, the EMI voltages of the Gent model are lower than those based on the neo-Hookean model. The voltage provided by the Gent model will first go up to the EMI voltage and then decline gradually with the increasing radial stretch, which is remarkably distinguished from the results for $\bar{N}=0$. This phenomenon is a consequence of the competition between the change in the electrostatic stress generated by the electric loading and that in the mechanical stress due to the strain-stiffening effect. Specifically, owing to the relatively low EMI voltage for $\bar{N}=0$, the generated electrostatic stress is small. When the DE cylinder reaches the strain-stiffening stage which can be captured by the Gent model, the mechanical stress will have a significant increase. Therefore, in order to make the radial stretch increase further, a higher voltage is required to produce a larger electrostatic stress to balance the induced mechanical stress, which leads to the rise of the applied voltage so as to exceed the EMI voltage. In contrast, the high EMI voltage for $\bar{N}=2.5$ and 5 will give rise to a larger electrostatic stress, which surpasses the increase of the mechanical stress even if the strain-stiffening effect appears. Consequently, the voltage should decline continuously for the purpose of ensuring the increase of the radial stretch for the axisymmetric deformation in the considered range of stretch in Figs. 11(b) and 11(c).

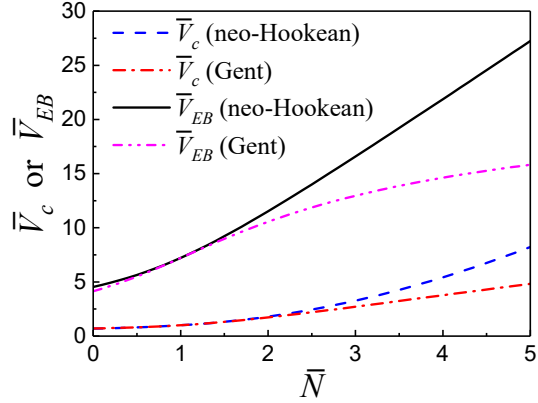

**Fig. 12.** Normalized critical voltages $\bar{V}_c$ corresponding to $\alpha = 1$ and the normalized EB voltages $\bar{V}_{EB}$ versus the axial force $\bar{N}$ in the DE phononic cylinder (Path B) for both neo-Hookean and Gent models.

Similar to loading path A, there also exists a critical electric voltage $\bar{V}_c$ corresponding to $\alpha = 1$ for loading path B. We plot the two voltages $\bar{V}_c$ and $\bar{V}_{EB}$ as functions of the axial force $\bar{N}$ of loading path B in Fig. 12 for both neo-Hookean and Gent models. Note that $\bar{V}_{EB} = E_{EB}\lambda_3\sqrt{\varepsilon/\mu}$ is calculated for the configuration subjected to the critical voltage $\bar{V}_c$. For prescribed axial force and electric voltage, the two material models give different axial stretches resulting in different EB voltages $\bar{V}_{EB}$, which differs from the result for loading path A. One notes that the critical voltage is lower than the EB voltage and increases monotonically with the axial force for either material model. Furthermore, through our numerical calculation, the critical voltages corresponding to $\bar{N} = 0, 2.5$ and 5 for both neo-Hookean and Gent models are equal to the respective values of the EMI voltage shown in Fig. 11, i.e., $\bar{V}_c = \bar{V}_{EMI}$. Therefore, when the applied voltage arrives at $\bar{V}_{EMI}$, the frequency limit (41)$_2$ will vanish, which will be displayed below. In brief, the allowable range of the applied voltage for loading path B is assumed to be $(0, \bar{V}_{EMI})$ for these two material models except for one case, in which the snap-through instability happens at $\bar{N} = 0$ for the Gent model. In addition, the critical voltages for both material models are almost the same for $\bar{N} \leq 2.5$, while the critical voltage predicted by the Gent model is smaller than that of the neo-Hookean model for $\bar{N} > 2.5$. This is opposite to the result for loading path A. This phenomenon can be easily understood because a larger extension induced by the same axial force for the neo-Hookean model shown in Fig. 3 needs a higher electric voltage to induce the EMI phenomenon.

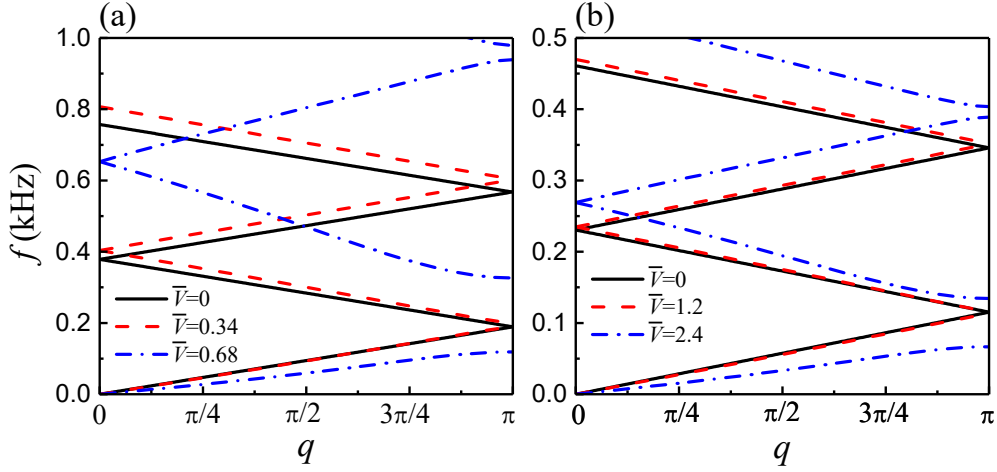

**Fig. 13.** Dispersion diagrams of the incremental longitudinal waves in the DE phononic cylinder (Path B) for the neo-Hookean model at different electric voltages and two different axial forces: (a) $\bar{N}=0$; (b) $\bar{N}=2.5$.

Fig. 13 shows the dispersion diagrams of the waves propagating in the DE phononic cylinder for the neo-Hookean model at different electric voltages lower than the EMI value for two different axial forces $\bar{N}=0$ and 2.5 of loading path B. Similar observations to those in Fig. 6 can be obtained although the axial force now plays the role instead of the axial pre-stretch. For the Gent model, qualitatively similar results can be achieved and hence are omitted here.

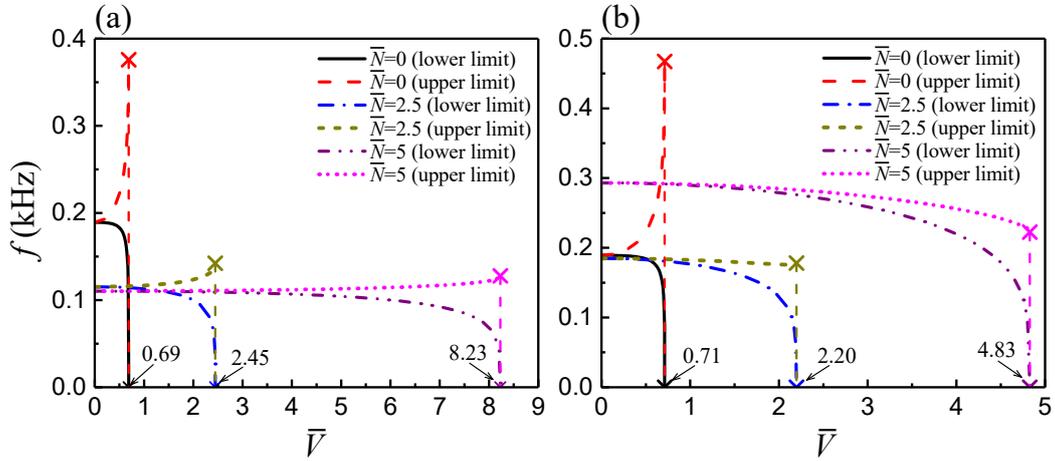

**Fig. 14.** Frequency limits of the first Bragg BG versus the dimensionless electric voltage $\bar{V}$ in the DE phononic cylinder (Path B) at different axial forces for both neo-Hookean (a) and Gent (b) models.

The frequency limits of the first Bragg BG versus the applied electric voltage in the DE phononic cylinder are depicted in Fig. 14 for both neo-Hookean and Gent models at three different axial forces $\bar{N}=0$, 2.5 and 5 of loading path B. Note that the snap-through instability at $\bar{N}=0$ for the Gent model are not yet taken into account in Fig. 14. It is found from Fig. 14 that, when the voltage increases to the

EMI voltage $\bar{V}_{EMI}$, the lower frequency limit of the first band gap will become zero for both material models. For the neo-Hookean model, comparing Fig. 14(a) with Fig. 7(a), we can obtain qualitatively analogous characteristics due to the axial force instead of the axial pre-stretch. For the Gent model displayed in Fig. 14(b), as the applied voltage increases from zero to the EMI voltage, the lower frequency limit decreases monotonically to zero and the upper frequency limit is elevated for $\bar{N}=0$. However, the lower and upper frequency limits for $\bar{N}=2.5$ and 5 all reduce with the increasing voltage. In fact, we first recall that the upper frequency limit is governed by Eq. (41)$_1$ with $\kappa^2 = \bar{\rho}/\bar{\mathcal{A}}_0^e$ and $\varpi = \lambda_3 \Omega$. When the applied voltage increases for $\bar{N}=0$, the effective elastic stiffness $\bar{\mathcal{A}}_0^e$ of the DE cylinder rises and the axial stretch decreases with the density $\bar{\rho}$ being kept almost unchanged, which results in a remarkable increase in frequency, as shown in Fig. 14(b). However, when the axial forces $\bar{N}=2.5$ and 5 are applied to the DE cylinder, the strain-stiffening stage has been reached where $\bar{\mathcal{A}}_0^e$ becomes relatively large. When the voltage goes up from zero to $\bar{V}_{EMI}$, the DE cylinder shortens but is still in a state of extension, which leads to a larger decrease in $\bar{\mathcal{A}}_0^e$ than that in the axial stretch $\lambda_3$. As a result, the upper frequency limits for both $\bar{N}=2.5$ and 5 have a drop. However, the width of the first BG still increases with the applied voltage for the Gent model.

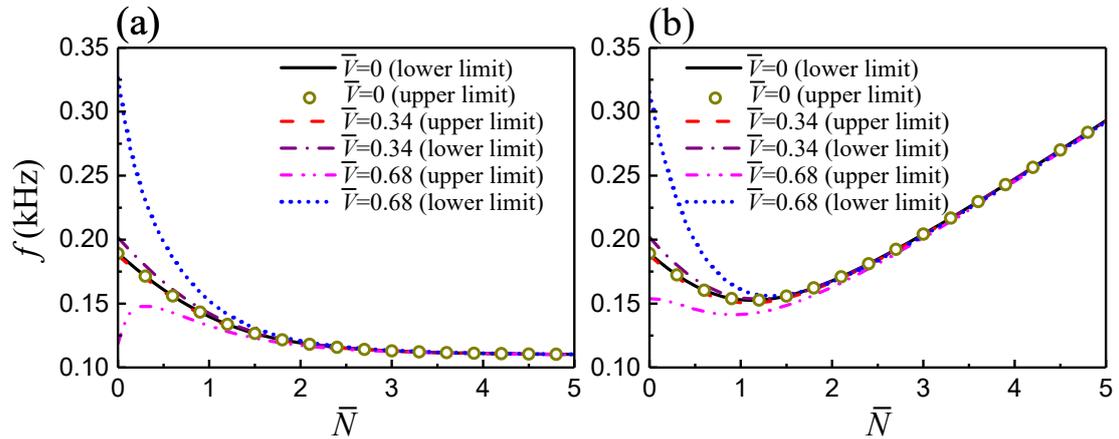

**Fig. 15.** Frequency limits of the first Bragg BG versus the axial force $\bar{N}$ in the DE phononic cylinder (Path B) at different electric voltages for both neo-Hookean (a) and Gent (b) models.

For loading path B, the variations of the frequency limits of the first Bragg BG with the axial force are presented in Fig. 15 for both neo-Hookean and Gent models at three different voltages lower than the EMI voltage. Making a comparison of Fig. 15(a) with Fig. 8(a), we can obtain essentially similar variation properties induced by either the axial pre-stretch or the axial force for the neo-Hookean model. Nonetheless, it is seen from Fig. 15(b) that, for the Gent model, all upper and lower frequency limits first decrease to a minimum and then increase continuously, which is quite different from the result for loading path A shown in Fig. 8(b).

Table 2. Main characteristics of the first tunable Bragg BG for the DE phononic cylinder at three different axial forces of both neo-Hookean (N-H) and Gent models (Path B). The first BG central frequency without the applied electric voltage is used for normalization and the snap-through instability at $\bar{N}=0$ for the Gent model is not yet taken into account.

| Axial force | Control range of normalized electric voltage ($\bar{V}$) | 1st band gap central frequency (kHz) | Variation of 1st band gap normalized central frequency | 1st band gap width (kHz) | Variation of 1st band gap normalized frequency width |
|---|---|---|---|---|---|
| $\bar{N}=0$ | N-H: 0-0.69 | 0.189-0.188 | 1-0.995 | 0-0.376 | 0-1.989 |
| | Gent: 0-0.71 | 0.189-0.234 | 1-1.236 | 0-0.467 | 0-2.471 |
| $\bar{N}=2.5$ | N-H: 0-2.45 | 0.115-0.071 | 1-0.618 | 0-0.142 | 0-1.235 |
| | Gent: 0-2.20 | 0.185-0.089 | 1-0.481 | 0-0.178 | 0-0.962 |
| $\bar{N}=5$ | N-H: 0-8.23 | 0.110-0.064 | 1-0.581 | 0-0.128 | 0-1.164 |
| | Gent: 0-4.83 | 0.293-0.111 | 1-0.380 | 0-0.222 | 0-0.758 |

Similar to Table 1, we conclude the main characteristics of the first tunable Bragg BG in Table 2 for both neo-Hookean and Gent models at three different axial forces of loading path B. Note that the snap-through instability at $\bar{N}=0$ for the Gent model is not included in Table 2, and will be discussed separately in the following. It is found that the variation ranges of the normalized central frequency and the normalized BG width are different for different axial forces and different material models. When compared to those reported for the hard piezoelectric phononic rod [52], analogous conclusions to those in Table 1 regarding the variation properties of the BGs can be made.

The nonlinear response of the axially free DE phononic cylinder with snap-through instability is again displayed in Fig. 16(a), which is the same as that in Fig. 11(a), but with two additional blue dashed arrows denoting the snap-through transitions. It should be emphasized that, through the numerical calculation, the DE phononic cylinder with high dielectric strength $E_{EB}=372$ MV/m survives the snap-through transition from state B to state C, reaches a stable state C before EB phenomenon, and achieves a large actuation deformation [55]. As previously mentioned, we can harness the snap-through instability to achieve sudden transitions in the BGs. In order to clearly show the characteristic, the variations of the frequency limits of the first Bragg BG with the applied voltage are illustrated in Fig. 16(b) for the axially free DE phononic cylinder based on the Gent model, with the snap-through transitions included. When the voltage increases from zero to $\bar{V}_{EMI} = 0.712$ (i.e., from state $A$ to state $B$ in Fig. 16(a)), a BG opens and its width increases from zero to 0.467kHz with its central frequency varying from 0.189kHz to 0.234kHz. This gap is indicated in Fig. 16(b) by the magenta region. A further incremental increase will make the snap-through transition happen and the DE cylinder snaps from state $B$ to state $C$ in Fig. 16(a), which results in sudden and enormous change in the BG width from 0.467kHz to 1.623kHz and that in the BG central frequency from 0.234kHz to 2.148kHz. A subsequent fall in voltage to $\bar{V} = 0.677$ (i.e., from state $C$ to state $D$ in Fig. 16(a)) hardly changes the BG width but indeed lowers the position of the BG with its central frequency altering from 2.148kHz to 0.745kHz. This process is denoted in Fig. 16(b) by the purple region. If the voltage has a further incremental decrease, a contracting snap-through transition from state $D$ to state $E$ is triggered, as shown by the left blue arrow in Fig. 16(a). This snap-through again leads to a sudden jump in the BG width from 1.489kHz to 0.156kHz and that in the BG central frequency from 0.745kHz to 0.234kHz. Successive decrease in voltage to zero will close the BG, as indicated in Fig. 16(b) by the blue region. In fact, the snap-through instabilities of thick-walled electroactive balloons have been exploited by Rudykh et al. [56] to electrically control different actuation cycles as well as to use the balloons as micro-pumps. In addition, the snap-through transitions have been utilized by Bortot and Shmuel [32] in the 2D sonic crystal consisting of an array of DE tubes in air to achieve sudden jumps in the BG width.

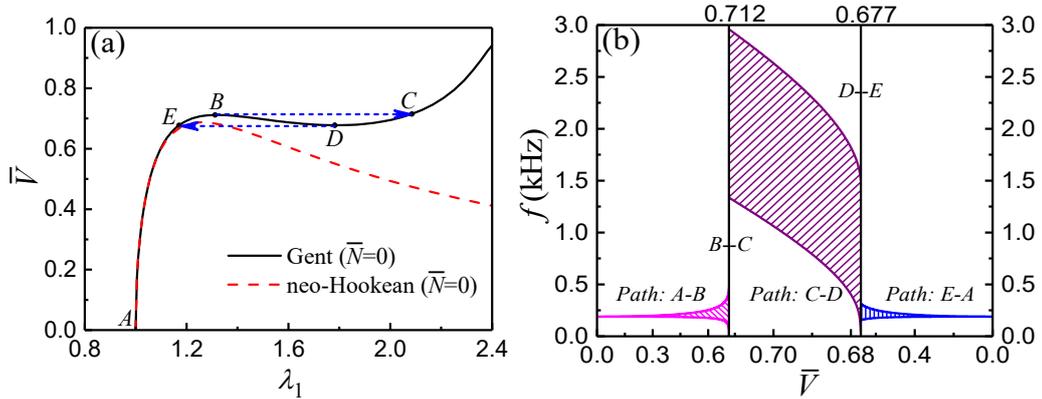

**Fig. 16.** (a) Nonlinear response of the radial stretch $\lambda_1$ to the dimensionless electric voltage $\bar{V}$ in the axially free DE phononic cylinder for the neo-Hookean and Gent models. The snap-through transitions associated with the Gent model only are denoted by the blue dashed arrows. (b) The frequency limits of the first Bragg BG versus the dimensionless electric voltage $\bar{V}$ in the axially free DE phononic cylinder for the Gent model.

## 7. Conclusions

In this paper, we investigated the longitudinal wave propagation in the DE phononic cylinder with periodic electrical boundary conditions. Firstly, two loading paths, i.e., fixed axial pre-stretch (Path A) and fixed axial force (Path B) were considered to form the effective periodicity of the soft PC cylinder and their nonlinear responses for both neo-Hookean and Gent models to the combined action of the applied voltage and axial force were able to be determined exactly without any approximation based on the nonlinear electroelasticity theory. Secondly, utilizing the assumption of one-dimensional stress state, the explicit dispersion relation for the superimposed longitudinal wave motions were derived approximately. Furthermore, the frequency limits of band gaps at the center and border of the first Brillouin zone and the long wave limits were achieved analytically. Finally, detailed numerical calculations were conducted to illustrate the static axisymmetric deformation response and the effects of the axial pre-stretch or axial force and the electric voltage on the band structure of the longitudinal waves for the two loading paths and the two nonlinear material models.

From the numerical results, we obtained the following primary observations: 1) there exist critical electric voltages resulting in zero frequency limit of the first BG and zero effective wave velocity for both loading paths; 2) the DE phononic cylinder with fixed axial force exhibits EMI phenomenon at the EMI voltage which is equal to the critical voltage; 3) a nonzero voltage opens the Bragg BGs at the border of the first Brillouin zone, and hence the DE phononic cylinder can be a proper candidate as an electrical switching device between non-dispersive waves and dispersive waves with Bragg BGs; 4) the applied voltage can largely widen the BGs while the axial pre-stretch or axial force mainly change their positions; 5) owing to the low frequency magnitude of the BGs, the DE phononic cylinder presents a potential application as low frequency filters; 6) the long wave limits agree well with the exact solution at low frequency and low wave number, and the effective wave velocity at long wave limits can be tuned by the biasing fields; 7) the axially free DE phononic cylinder described by the Gent model presents the snap-through instability resulting from geometrical and material nonlinearities, which can be exploited to realize sharp transitions in the BGs. All these results indicate that both large deformation and electromechanical coupling play an important role in the longitudinal wave propagation in the DE PCs. Consequently, they can be efficient means to engineer the band structure, and in particular to tune the BGs in soft electroactive PCs.

Note that the first Lamé's parameter and the Gent constant in this analysis have been fixed. Further investigation on the influence of these two parameters upon the longitudinal wave propagation characteristics (e.g., the band structure) should be interesting. In addition, when slender soft phononic rods are subjected to compressive axial force (or pre-stretch), they will be susceptible to buckling instabilities. Therefore, the effect of compressive axial force on the band structure, which has not been considered in this paper will be more complex, but it could be an interesting topic of future research.

One recent interesting work accomplished by Zhang and Parnell [57] is finally noticed here. Actually, they initiated an investigation opposite to tunable PCs, i.e. they studied how the band structure in a soft PC could be kept unchanged during the large deformation. This is extremely important when we expect to design a PC that can work robustly with a consistent performance. When electromechanical coupling and complex geometry present, the problem becomes more challenging and deserves further study.

**Acknowledgements**

This work was supported by the National Natural Science Foundation of China (Nos. 11532001 and 11621062). Partial support from the Fundamental Research Funds for the Central Universities (No. 2016XZZX001-05) is also acknowledged.

**Appendix: Some analytical results for the neo-Hookean model**

In this appendix, we give some analytical results for the ideal compressible neo-Hookean model that are useful for performing the numerical calculation. The purely elastic part of the energy density function (11) for compressible neo-Hookean materials can be obtained from Eq. (12) by setting $J_m \to \infty$ as

$$W_{elas}^{nH} = \frac{\mu}{2}(I_1 - 3) - \mu \ln J + \frac{\Lambda}{2}(J-1)^2 \tag{A1}$$

In the limit of $J_m \to \infty$, Eqs. (13), (17), (19)-(21) and (27), which are for the Gent model, become those for the neo-Hookean model as follows:

$$\sigma_1 = \sigma_2 = \mu J^{-1}\left(\lambda_1^2 - 1\right) + \Lambda(J-1) - \frac{D_3^2}{2\varepsilon},$$
$$\sigma_3 = \mu J^{-1}\left(\lambda_3^2 - 1\right) + \Lambda(J-1) + \frac{D_3^2}{2\varepsilon}, \quad E_3 = \frac{D_3}{\varepsilon} \tag{A2}$$

$$\bar{\sigma}_1 = \frac{1}{J^{pre}}\left[\left(\lambda_1^{pre}\right)^2 - 1\right] + \bar{\Lambda}(J^{pre} - 1) = 0,$$
$$\bar{\sigma}_3 = \frac{1}{J^{pre}}\left[\left(\lambda_3^{pre}\right)^2 - 1\right] + \bar{\Lambda}(J^{pre} - 1) = \frac{\bar{N}}{\left(\lambda_1^{pre}\right)^2} \tag{A3}$$

$$\bar{\sigma}_1 = \frac{1}{J}\left(\lambda_1^2 - 1\right) + \bar{\Lambda}(J-1) - \frac{\bar{V}^2}{2\lambda_3^2} = 0 \tag{A4}$$

$$\bar{\sigma}_3 = \frac{1}{J}\left[(\lambda_3)^2 - 1\right] + \bar{\Lambda}(J-1) + \frac{\bar{V}^2}{2(\lambda_3)^2}, \quad \bar{N} = \bar{\sigma}_3 \lambda_1^2 \tag{A5}$$

$$\begin{aligned}\bar{\sigma}_1 &= \frac{1}{J}\left(\lambda_1^2 - 1\right) + \bar{\Lambda}(J-1) - \frac{\bar{V}^2}{2\lambda_3^2} = 0, \\ \bar{\sigma}_3 &= \frac{1}{J}\left(\lambda_3^2 - 1\right) + \bar{\Lambda}(J-1) + \frac{\bar{V}^2}{2\lambda_3^2} = \frac{\bar{N}}{\lambda_1^2}\end{aligned} \tag{A6}$$

$$\begin{aligned}\bar{\mathcal{A}}_{01111} &= \bar{\mathcal{A}}_{02222} = \frac{1}{J}\left(\lambda_1^2 + 1 + J^2\bar{\Lambda} + J\bar{D}_3^2\right), \quad \bar{\mathcal{A}}_{01122} = \bar{\Lambda}(2J-1) + \frac{\bar{D}_3^2}{2}, \\ \bar{\mathcal{A}}_{01133} &= \bar{\mathcal{A}}_{02233} = \bar{\Lambda}(2J-1) - \frac{\bar{D}_3^2}{2}, \quad \bar{\mathcal{A}}_{03333} = \frac{1}{J}\left(\lambda_3^2 + 1 + J^2\bar{\Lambda}\right)\end{aligned} \tag{A7}$$